\input harvmac.tex

%%%%%%%%%%%%% FOR THE FIGURES
%%%%%%%%%%%%%
\input epsf.tex
\def\figin{\epsfcheck\figin}\def\figins{\epsfcheck\figins}
\def\epsfcheck{\ifx\epsfbox\UnDeFiNeD
\message{(NO epsf.tex, FIGURES WILL BE IGNORED)}
\gdef\figin##1{\vskip2in}\gdef\figins##1{\hskip.5in}% blank space instead
\else\message{(FIGURES WILL BE INCLUDED)}%
\gdef\figin##1{##1}\gdef\figins##1{##1}\fi}
\def\DefWarn#1{}
\def\figinsert{\goodbreak\midinsert}
\def\ifig#1#2#3{\DefWarn#1\xdef#1{fig.~\the\figno}
\writedef{#1\leftbracket fig.\noexpand~\the\figno}%
\figinsert\figin{\centerline{#3}}\medskip\centerline{\vbox{\baselineskip12pt
\advance\hsize by -1truein\noindent\footnotefont{\bf Fig.~\the\figno:} #2}}
\bigskip\endinsert\global\advance\figno by1}
%%%%%%%%%%%%%
%%%%%%%%%%%%%

\def\p{\partial}
\def\mod{{\rm mod}}

\def\CP{{\cal P}}
\def\ndt{\noindent}
\def\bgn{\bigskip\ndt}

%% MACROS

\def\IL{\relax{\rm I\kern-.18em L}}
\def\IH{\relax{\rm I\kern-.18em H}}
\def\IR{\relax{\rm I\kern-.18em R}}
\def\IC{\relax\hbox{$\inbar\kern-.3em{\rm C}$}}

\def\CR {{\cal R}}

\def\CP {{\cal P }}

\def\CO {{\cal O}}

\def\CS {{\cal S}}

\def\CW{{\cal W}}
\def\IZ{{\bf Z}}
%% MORE MACROS

\def\CO {{\cal O}}

\def\CP {{\cal P }}
\def\CQ {{\cal Q }}

\def\CS {{\cal S }}

\font\manual=manfnt \def\dbend{\lower3.5pt\hbox{\manual\char127}}

\def\half {{1\over 2}}

\def\p{\partial}

\def\vol{{\rm vol}}

%%%%%%%%%%%%%%%%%%%%%%
%%%%%%%%%%%%%%%%%%%%%% REFS

%
\lref\ars{A. Alekseev, A. Recknagel, and V. Schomerus,
``Non-commutative World-volume Geometries: Branes on SU(2) and
Fuzzy Spheres,''hep-th/9908040; ``Brane Dynamics in Background
Fluxes and Non-commutative Geometry,'' hep-th/0003187}

\lref\asii{A. Alekseev and V. Schomerus,
``D-branes in the WZW model,'' hep-th/9812193;
Phys. Rev. {\bf D60}(1999)061901}
\lref\as{A. Alekseev and V. Schomerus,
 ``RR charges of D2-branes in the WZW model,''
hep-th/0007096}

\lref\atiyahhirz{M.F. Atiyah and F. Hirzebruch, ``Vector Bundles
and Homogeneous Spaces,'' Proc. Symp. Pure Math. {\bf 3} (1961) 53.}

\lref\emssr{ S.~Elitzur, G.~Moore, A.~Schwimmer and N.~Seiberg,
``Remarks On The Canonical Quantization Of The
Chern-Simons-Witten Theory,'' Nucl.\ Phys.\ B {\bf 326}, 108
(1989).
%%CITATION = NUPHA,B326,108;%%
}
\lref\atiyahbook{M. Atiyah, {\it K-theory}, Addison-Wesley, 1989. }

\lref\atiyahsegal{M. Atiyah and G.B. Segal, unpublished.}

\lref\bds{C.~Bachas, M.~Douglas and C.~Schweigert,
``Flux stabilization of D-branes,''
JHEP {\bf 0005}, 048 (2000)
[hep-th/0003037].
%%CITATION = HEP-TH 0003037;%%
}

\lref\birke{L. Birke, J. Fuchs, and C. Schweigert, ``Symmetry
breaking boundary conditions and WZW orbifolds,'' Adv.\ Theor.\
Math.\ Phys.\ { 3}, 671 (1999);  hep-th/9905038}
\lref\bm{P. Bouwknegt and V. Mathai,  ``$D$-Branes, $B$ Fields
and Twisted $K$-Theory,'' JHEP {\bf 0003:007,2000}, hep-th/0002023.
}
\lref\cvetic{M. Cvetic, H. Lu, C.N. Pope, ``Consistent Kaluza-Klein
sphere reductions,'' hep-th/0003286;
M. Cvetic, H. Lu, C.N. Pope, A. Sadrzadeh, and
T.A. Tran, ``$S^3$ and $S^4$ reductions of type IIA supergravity,''
hep-th/0005137}
\lref\cftbook{P, Di Francesco, P. Mathieu,
D. S\'en\'egal, {\it Conformal Field
Theory} Springer, 1997
}

\lref\dmw{E. Diaconescu, G. Moore, and E. Witten, ``$E_8$ Gauge Theory,
and a Derivation of $K$-Theory from $M$-Theory,'' hep-th/0005090}

\lref\douglasnc{M. Douglas, ``D-Branes in Curved Space,'' hep-th/9703056;
``D-branes and Matrix Theory in Curved Space,''hep-th/9707228;
``Two Lectures on D-Geometry and Noncommutative Geometry,''
hep-th/9901146.}

\lref\duff{M. Duff et. al. }
\lref\epstein{Epstein and N. Steenrod, {\it Cohomology Operations},
Ann. Math. Stud. {\bf 50}, Princeton Univ. Press}

\lref\fffs{G. Felder, J. Fr\"ohlich, J. Fuchs, C. Schweigert,
``The geometry of WZW branes,''hep-th/9909030,
J.Geom.Phys. 34 (2000) 162-190}

\lref\fuchs{J. Fuchs, {\it Affine Lie Algebras and Quantum Groups},
Cambridge Univ. Press}

\lref\figeroa{J.M. Figeroa-O'Farrill and S. Stanciu, ``D-brane charge,
flux quantization and relative (co)homology,'' hep-th/0008038}
\lref\fredenhagen{S. Fredenhagen and V. Schomerus,
``Branes on Group Manifolds, Gluon Condensates, and twisted K-theory,''
hep-th/0012164}

\lref\freedwitten{D.S. Freed and E. Witten, ``Anomalies in String Theory
with $D$-branes,'' hep-th/9907189.}
\lref\freedhopkins{D.S. Freed and M.J. Hopkins,
``On Ramond-Ramond fields and K-theory,'' hep-th/0002027 JHEP 0005 (2000) 044}
\lref\freed{D. Freed,``Dirac Charge Quantization and Generalized
Differential Cohomology,''
  hep-th/0011220}
\lref\gawedzki{K. Gawedzki, ``Conformal field theory: a case study,''
hep-th/9904145}
\lref\gibbons{G.W. Gibbons and K.  Maeda,
``Black holes and membranes in higher dimensional theories
with dilaton fields,''
 Nucl.Phys.B298:741,1988 }

\lref\unwinding{R. Gregory, J. Harvey, and G. Moore,
``Unwinding strings and T-duality of Kaluza-Klein and H-Monopoles,''
hep-th/970808; Adv.Theor.Math.Phys. 1 (1997) 283-297}
\lref\hm{J. Harvey and G. Moore, ``Noncommutative Tachyons and K-Theory,''
hep-th/0009030.}
\lref\hs{G.~T.~Horowitz and A.~Strominger,
``Black strings and P-branes,''
Nucl.\ Phys.\ B {\bf 360}, 197 (1991).
%%CITATION = NUPHA,B360,197;%%
}

\lref\kapustin{A.~Kapustin,
``D-branes in a topologically nontrivial B-field,''
Adv.\ Theor.\ Math.\ Phys.\  {\bf 4}, 127 (2001)
[hep-th/9909089].
%%CITATION = HEP-TH 9909089;%%
}

\lref\larsenmartinec{F.~Larsen and E.~Martinec,
``U(1) charges and moduli in the D1-D5 system,''
JHEP {\bf 9906}, 019 (1999)
[hep-th/9905064].
%%CITATION = HEP-TH 9905064;%%
}

\lref\mmsi{J. Maldacena, G. Moore, and N. Seiberg, Parafermions}

\lref\mmsii{J. Maldacena, G. Moore, and N. Seiberg, ``D-brane
instantons and K-theory charges,'' hep-th/0108100.}

\lref\mathaisinger{V. Mathai and I.M. Singer, ``Twisted
K-homology theory, twisted Ext-theory,'' hep-th/0012046}

\lref\marolf{D. Marolf, ``Chern-Simons terms and the three
notions of charge,'' hep-th/0006117}
\lref\mm{R. Minasian and G. Moore,``K-Theory and Ramond-Ramond
Charge,'' JHEP {\bf 9711} :002, 1997; hep-th/9710230.}

\lref\morozov{A. Alekseev, A. Mironov, and A. Morozov,
``On B-independence of RR charges,'' hep-th/0005244}
\lref\polyakov{A.M. Polyakov, PLB59 (1975)82;
``Quark confinement and topology of gauge fields,''
Nucl. Phys. {\bf B120} (1977) 429 ; {\it Gauge Fields
and Strings} Harwood Academic Publishers, 1987, ch. 4}
\lref\selfduality{G. Moore and E. Witten, ``Self-duality, RR fields
and $K$-Theory,'' hep-th/9912279.}

\lref\rosenberg{ J. Rosenberg, ``Homological Invariants of
Extensions of $C^*$-algebras,'' Proc. Symp. Pure Math {\bf 38}
(1982) 35.}

\lref\stanciu{S. Stanciu, ``D-branes in group manifolds,''
hep-th/9909163}
\lref\standiuii{S. Stanciu, ``A note on D-branes in
group manifolds: flux quantization and D0 charge,''
hep-th/0006145}
\lref\taylor{W.~Taylor,
``D2-branes in B fields,''
JHEP {\bf 0007}, 039 (2000)
[hep-th/0004141].
%%CITATION = HEP-TH 0004141;%%
}

\lref\townsend{P.~K.~Townsend, ``D-branes from M-branes,'' Phys.\
Lett.\ B {\bf 373}, 68 (1996) [hep-th/9512062].
%%CITATION = HEP-TH 9512062;%%
}

\lref\wendt{R. Wendt, ``Weyl's character formula for non-connected
Lie groups and orbital theory for twisted affine Lie algebras,''
math.RT/9909059}
\lref\wittenjones{E. Witten, ``Quantum field theory and the Jones
polynomial,'' Commun.Math.Phys. 121:351, 1989}

\lref\wittenbaryon{E. Witten
``Baryons And Branes In Anti de Sitter Space,''
hep-th/9805112;JHEP 9807 (1998) 006}
\lref\wittenk{E. Witten, ``$D$-Branes And $K$-Theory,''
JHEP {\bf 9812}:019, 1998; hep-th/9810188.}
\lref\wittenstrings{E. Witten, ``Overview of K-theory applied to
strings,'' hep-th/0007175.}
%

%%%%%%%%%%%%%%%%%% NEW REFERENCES %%%%%%%%%%%%%%%%%%%%%%

\lref\polchinski{ J.  Polchinski, ``String theory'',
Cambridge Univ. Press, 1998. }

\lref\itzhaki{
N.~Itzhaki, J.~M.~Maldacena, J.~Sonnenschein and S.~Yankielowicz,
``Supergravity and the large N
limit of theories with sixteen  supercharges,''
Phys.\ Rev.\ D {\bf 58}, 046004 (1998)
[hep-th/9802042].}

\lref\mn{J.~M.~Maldacena and C.~Nunez,
``Towards the large n limit of pure N = 1 super Yang Mills,''
Phys.\ Rev.\ Lett.\  {\bf 86}, 588 (2001)
[hep-th/0008001].
}

\lref\superpotential{J. A. Harvey and G. Moore,
``Superpotentials and Membrane Instantons,''
hep-th/9907026;
G.~Moore, G.~Peradze and N.~Saulina,
``Instabilities in
heterotic M-theory induced by open membrane  instantons,''
hep-th/0012104; E.~Lima, B.~Ovrut and J.~Park,
``Five-brane superpotentials in heterotic M-theory,''
hep-th/0102046.}

\lref\zhou{J.~Zhou,
``Page charge of D-branes
and its behavior in topologically nontrivial B-fields,''
hep-th/0105106.
}

\lref\gukov{
S.~Gukov,
``K-theory, reality, and orientifolds,''
Commun.\ Math.\ Phys.\  {\bf 210}, 621 (2000)
[hep-th/9901042].
%%CITATION = HEP-TH 9901042;%%
}

\lref\PAH{ J.~Pawelczyk, ``SU(2) WZW D-branes and their
noncommutative geometry from DBI action,'' JHEP {\bf 0008}, 006
(2000) [hep-th/0003057]; J.~Pawelczyk and S.~Rey, ``Ramond-Ramond
flux stabilization of D-branes,'' Phys.\ Lett.\ B {\bf 493}, 395
(2000) [hep-th/0007154].
%%CITATION = HEP-TH 0007154;%%
}

\lref\sencc{A. Sen, ``Kaluza-Klein Dyons in String Theory,''
hep-th/9705212, Phys.Rev.Lett. 79 (1997) 1619-1621; ``Dynamics of
Multiple Kaluza-Klein Monopoles in M- and String Theory,''
hep-th/9707042, Adv.Theor.Math.Phys. 1 (1998) 115-126}

\lref\fourflux{E. Witten, ``On Flux Quantization in $M$-Theory
and the Effective Action,'' hep-th/9609122; Journal of
Geometry and Physics, {\bf 22} (1997) 1.}

\lref\wittenadstft{E. Witten, ``AdS/CFT Correspondence and
Topological Field Theory'',\hfill\break JHEP {\bf 9812} (1998)
012, hep-th/98120112.}

\lref\imwitten{E. Witten, ``Five-Brane Effective Action In
$M$-Theory'', J. Geom. Phys. {\bf 22} (1997) 103, hep-th/9610234.}

\lref\baryons{E. Witten, ``Baryons and Branes in Anti de Sitter
Space, JHEP {\bf 9807} (1998) 006, hep-th/9805112.}

\lref\duality{E.~Witten, ``Duality relations among topological
effects in string theory,'' JHEP {\bf 0005}, 031 (2000)
[hep-th/9912086].
%%CITATION = HEP-TH 9912086;%%
}

\lref\KlebanovHB{ I.~R.~Klebanov and M.~J.~Strassler,
``Supergravity and a confining gauge theory: Duality cascades
and  chiSB-resolution of naked singularities,'' JHEP {\bf 0008},
052 (2000) [hep-th/0007191].
%%CITATION = HEP-TH 0007191;%%
}

\lref\cheegersimons{J. Cheeger and J. Simons, ``Differential Characters
and Geometric Invariants,'' in {\it Geometry and Topology}, J. Alexander
and J. Harer eds., LNM 1167.}

\lref\harris{B. Harris, ``Differential Characters and the Abel-Jacobi
Map'', in {\it Algebraic $K$-theory: Connections with Geometry and
Topology},
J.F. Jardine and V.P. Snaith eds., Nato ASI Series C: Mathematical
and Physical Sciences  -- Vol. 279, Kluwer Academic Publishers,
1989.}

\lref\dijkgraafwitten{R. Dijkgraaf and E. Witten,
``Topological gauge theories and group cohomology,''
  Commun.Math.Phys.129:393,1990 }
\lref\gawedzki{K. Gawedzki, ``Topological actions in two-dimensional
quantum field theories,'' in {\it Nonperturbative quantum field theory}
Procedings of the 1987 Cargese meeting, pp. 101-141.}
\lref\zucchini{R. Zucchini, ``Relative topological integrals and
relative Cheeger-Simons differential characters,'' hep-th/0010110}

\lref\dmunpub{E. Diaconescu and G. Moore, unpublished.}

\lref\wittenhol{
E.~Witten,
``Anti-de Sitter space and holography,''
Adv.\ Theor.\ Math.\ Phys.\  {\bf 2}, 253 (1998)
[hep-th/9802150].
%%CITATION = HEP-TH 9802150;%%
}

\lref\aw{
O.~Aharony and E.~Witten,
``Anti-de Sitter space and the center of the gauge group,''
JHEP {\bf 9811}, 018 (1998)
[hep-th/9807205].
%%CITATION = HEP-TH 9807205;%%
}

\lref\klebanovwitten{
I.~R.~Klebanov and E.~Witten,
``Superconformal field theory on threebranes
at a Calabi-Yau  singularity,''
Nucl.\ Phys.\ B {\bf 536}, 199 (1998)
[hep-th/9807080].
%%CITATION = HEP-TH 9807080;%%
}

\lref\schse{ J.~H.~Schwarz and A.~Sen, ``Duality symmetric
actions,'' Nucl.\ Phys.\ B {\bf 411}, 35 (1994) [hep-th/9304154].
%%CITATION = HEP-TH 9304154;%%
}

\lref\MaldacenaBW{J.~Maldacena and A.~Strominger, ``AdS(3) black
holes and a stringy exclusion principle,'' JHEP {\bf 9812}, 005
(1998) [hep-th/9804085].
%%CITATION = HEP-TH 9804085;%%
}

\lref\MaldacenaRE{ J.~Maldacena, ``The large N limit of
superconformal field theories and supergravity,'' Adv.\ Theor.\
Math.\ Phys.\  {\bf 2}, 231 (1998) [Int.\ J.\ Theor.\ Phys.\
{\bf 38}, 1113 (1998)] [hep-th/9711200].
%%CITATION = HEP-TH 9711200;%%
}

\lref\GubserBC{ S.~S.~Gubser, I.~R.~Klebanov and A.~M.~Polyakov,
``Gauge theory correlators from non-critical string theory,''
Phys.\ Lett.\ B {\bf 428}, 105 (1998) [hep-th/9802109].
%%CITATION = HEP-TH 9802109;%%
}

\lref\WittenQJ{ E.~Witten, ``Anti-de Sitter space and
holography,'' Adv.\ Theor.\ Math.\ Phys.\  {\bf 2}, 253 (1998)
[hep-th/9802150].
%%CITATION = HEP-TH 9802150;%%
}

\lref\AharonyTI{ O.~Aharony, S.~S.~Gubser, J.~Maldacena,
H.~Ooguri and Y.~Oz, ``Large N field theories, string theory and
gravity,'' Phys.\ Rept.\  {\bf 323}, 183 (2000) [hep-th/9905111].
%%CITATION = HEP-TH 9905111;%%
}

\lref\SeibergXZ{ N.~Seiberg and E.~Witten, ``The D1/D5 system and
singular CFT,'' JHEP {\bf 9904}, 017 (1999) [hep-th/9903224].
%%CITATION = HEP-TH 9903224;%%
}

\lref\DijkgraafGF{ R.~Dijkgraaf, ``Instanton strings and
hyperKaehler geometry,'' Nucl.\ Phys.\ B {\bf 543}, 545 (1999)
[hep-th/9810210].
%%CITATION = HEP-TH 9810210;%%
}

\lref\KutasovXU{ D.~Kutasov and N.~Seiberg, ``More comments on
string theory on AdS(3),'' JHEP {\bf 9904}, 008 (1999)
[hep-th/9903219].
%%CITATION = HEP-TH 9903219;%%
}

\lref\MaldacenaBP{ J.~Maldacena, G.~Moore and A.~Strominger,
``Counting BPS black holes in toroidal type II string theory,''
hep-th/9903163.
%%CITATION = HEP-TH 9903163;%%
}

\lref\review{
O.~Aharony, S.~S.~Gubser, J.~Maldacena, H.~Ooguri and Y.~Oz,
``Large N field theories, string theory and gravity,''
Phys.\ Rept.\  {\bf 323}, 183 (2000)
[hep-th/9905111].
%%CITATION = HEP-TH 9905111;%%
}

\lref\mv{ M. Marino and  C. Vafa,
``Framed knots at large N''
hep-th/0108064.
}

\lref\BilalTY{ A.~Bilal and C.~Chu, ``Testing the AdS/CFT
correspondence beyond large N,'' hep-th/0003129.
%%CITATION = HEP-TH 0003129;%%
}

\lref\gk{
S.~S.~Gubser and I.~R.~Klebanov,
``Baryons and domain walls in an N = 1 superconformal gauge theory,''
Phys.\ Rev.\ D {\bf 58}, 125025 (1998)
[hep-th/9808075].
%%CITATION = HEP-TH 9808075;%%
}

\lref\maldaloop{J. Maldacena, ``Wilson loops in large N field
theorieds,'' Phys. Rev. Lett. {\bf 80}(1998) 4859; hep-th/9803002
}

\lref\reyloop{S.-J. Rey and J. Yee, ``Macroscopic strings as heavy quarks
in large N gauge theories and Anti-de Sitter supergravity,'' hep-th/9803001}

\lref\lott{J. Lott, ``R/Z Index Theory,'' Comm. Anal. Geom. 2, p. 279}

\lref\hopkinssinger{M. Hopkins and I.M. Singer, unpublished, but to appear. }

%%%%%%%%%%%%%%%%%%%%%%  END NEW REFERENCES %%%%%%%%%%%%%%%%%5
%%%%%%%%%%%%%%%%%%%%%%

%%%%%%%%%%%%%%%%%%%%%%
%%%%%%%%%%%%%%%%%%%%%%

%%%%%%%%%%%%%%%%%%%%%%%%%%%%%%%%%%%%%%%%%%%%%%%%%%%%%%%%%%%%%%%%%%%%%%%%%%
%%%%%%%%%%%%%%%%%%%%%%%%%%%%%%%%%%%%%%%%%%%%%%%%%%%%%%%%%%%%%%%%%%%%%%%%%%
\Title{\vbox{\baselineskip12pt
\hbox{hep-th/0108152}
\hbox{RUNHETC-2001-25 }
}}%
{\vbox{\centerline{D-brane charges  }
\medskip
\centerline{in  }
\medskip
\centerline{Five-brane backgrounds }
}}

\smallskip
\centerline{Juan Maldacena$^{1,2}$, Gregory Moore$^{3}$, Nathan
Seiberg$^{1}$ }
\medskip

\centerline{\it $^{1}$ School of Natural Sciences,}
\centerline{\it Institute for Advanced Study} \centerline{\it
Einstein Drive}
\centerline{\it Princeton, New Jersey, 08540}

\centerline{\it $^{2}$ Department of Physics, Harvard University}
\centerline{\it Cambridge, MA 02138}

\centerline{\it $^{3}$ Department of Physics, Rutgers University}
\centerline{\it Piscataway, New Jersey, 08855-0849}

\bigskip
\noindent
We discuss the discrete $\IZ_k$ D-brane charges (twisted
K-theory charges) in five-brane backgrounds from several
different points of view.  In particular, we interpret it as a
result of a standard Higgs mechanism. We show that certain
degrees of freedom (singletons) on the boundary of space can
extend the corresponding $\IZ_k$ symmetry to $U(1)$.  Related
ideas clarify the role of AdS singletons in the AdS/CFT
correspondence.

\medskip

\Date{August 14, 2001}

%\draftmode

%%%%%%%%%%%%%%%%%%%%%%%%%%%%%%%%%%%%%%%%%%%%%%%%%%%%%%%%%%%%%%%%%%%%%%%%%%
%%%%%%%%%%%%%%%%%%%%%%%%%%%%%%%%%%%%%%%%%%%%%%%%%%%%%%%%%%%%%%%%%%%%%%%%%%

\newsec{Introduction }

D-brane charges and RR fluxes are classified by K-theory, at weak
coupling \refs{\mm,\wittenk,\selfduality}. It is important to bear
in mind that all the arguments in favor of this rely on a picture
of type II strings valid at weak coupling and on smooth spaces.
Moreover, there is some tension between the K-theoretic
classification of charges and fluxes and both U-duality (which
mixes RR and NSNS degrees of freedom), and with the AdS/CFT
correspondence (which treats charges and fluxes more
democratically).

An important open problem is finding a more broadly applicable
homotopy classification of both charges with fluxes. Even
stating this open problem in a precise and useful way would
constitute some progress.

With this motivation in mind the present paper studies the
physical meaning of the K-theoretic  charge group $K_H^*(X)$ in
the presence of nontorsion $H$-flux, more specifically in
backgrounds associated with 5-branes\foot{It is an interesting
question whether any background with cohomologically nontrivial
H-flux can be regarded as one created by 5-branes. }. We will
find that in some situations the K-theoretic classification can
be slightly misleading, and we will show (in our examples) how to correct
it.

One of the hallmarks of $K$-theoretic charges
is that they naturally include torsion charges,
and this is often cited as an important distinction from
more traditional viewpoints on charges.
Nevertheless, one should not lose sight of the fact
that torsion charges also can arise naturally in standard
gauge theories. It is true that the   charges of
a $U(1)$ gauge group label the different representations and take
values in $Z$. But, if the
$U(1)$ gauge symmetry is broken to the integers
modulo $k$, denoted here by  $Z _k$,  through a standard
Higgs mechanism via a charged scalar field of charge $k>1$,
then the unbroken gauge symmetry is $Z_k$. The dual group of
representations is again $Z_k$ leading to a torsion group of
charges.  We will see that this is precisely what happens in
some examples of K-theoretic torsion charges.    We will also see in these
examples that there are often physical modes, zero-modes of
RR potentials,
which effectively restore the
$U(1)$ symmetry. In the context of string theory on AdS spaces
these modes reside at the boundary and
are sometimes called ``singletons'' (or ``doubletons'').
We will extend the terminology to a wider context and
refer to these boundary modes as ``singletons.''

In appendix A we discuss general properties of $p$-forms with
Chern Simons couplings and in appendix B we explain how the singleton
is related to the $U(1)$ subgroup of the gauge group  in
the AdS/CFT correspondence.

\subsec{The general class of backgrounds}

We will now be more precise about the class of backgrounds under
consideration.
We will focus on type IIA string theory with the action
\eqn\iialag{
\eqalign{
S=&
{1 \over (2\pi)^7 \alpha'^4} \int  \sqrt{-\det g } e^{-2\phi}
\bigl( \CR + 4 (\nabla \phi)^2\bigr)
 -{1\over 2 (2\pi)^3 \alpha'^2  } \int  e^{-2\phi}
H\wedge * H \cr
 &-{1\over 2  (2\pi)^7 \alpha'^3}\int R_2\wedge * R_2
 - {1 \over 2(2\pi)^3 \alpha' } \int R_4 \wedge * R_4
+ {1\over 4\pi} \int C_3 \wedge d C_3\wedge H
}
}
Here  $g$ is the string metric,
$\phi$ is the dilaton, $R_2:=dC_1$ is the RR 2-form fieldstrength,
and $R_4:=dC_3 - H\wedge C_1$.
\foot{When the M-theory circle is
a non-trivial fibration, or
  $C_3$ is not globally well-defined, or there is a Romans mass
then the above expression must be modified.}
\foot{The topological terms in this action and in the M-theory
action lead to a natural normalization convention for differential
forms: The fieldstrength of the 4-form flux in $M$-theory
 should be dimensionless and should have
periods which are $2\pi$ times an integral class (more precisely,
an integral class plus ${1\over 4} p_1(TX)$ \fourflux). This fixes
most of the conventions for the IIA Lagrangian. In particular
the $H$-flux has integral periods. In other words,
our normalizations
for RR fields differ from \polchinski\ in the following way: $
C_{p+1}^{us} = \mu_p C_{p+1}^{Polchinski} $, where $\mu_p^{-2} = (
4 \pi^2 \alpha')^{p} \alpha' $. Similarly, for the NS three form
we have  $(2 \pi)^2 \alpha'  H_3^{us} = H_3^{Polchinski}$.  }

We take spacetime to be a product of time and
9-dimensional space $\IR_t \times X_9$, where $X_9$ can be compact or
noncompact. Let us assume we have a IIA geometry with a {\it
static} metric on $\IR_t \times X_9$ such that near infinity we
have $X_9 = \IR \times X_8$ ($\IR$ is the radial direction), where
$X_8$ is fibered by $S^3\to X_8\to  X_5$. Moreover, we assume
that $S^3$ is not a contractible cycle in $X_9$.
 The NS three form fieldstrength $H=k \xi_3$ for some
3-form $\xi_3$ with integral periods such that $\xi_3 \to
\omega_3 (1 + \CO(1/r)) $ at infinity, where $\omega_3$ is
the volume form of $S^3$, normalized to have period $=1$.
The Bianchi identity and
the equation of motion imply that $d\xi_3 =0$ and $d(*e^{-2\phi}
\xi_3)=0$. At infinity $H \to k \omega_3$.

 The quintessential background with cohomologically
nontrivial $H$-flux is the NS5-brane of Callan, Harvey and
Strominger:
\eqn\iiafive{
\eqalign{
ds^2_{IIA} & = (dx)^2 + U (dy_{\bot})^2 \cr
e^{2\phi-2\phi_{\infty} } & = U = 1 + {k\alpha' \over  r^2 } \cr
H & = -k \omega_3 \cr}
}
where $\omega_3$ is the integral class on $S^3$
and $r:= \vert y_{\bot}\vert$. We will sometimes restrict
attention
to   the throat region
$e^{2\phi_{\infty}} \alpha'  \ll r^2 \ll k\alpha' $.
%
%Then  we can define a
%length-scale    $\xi := \sqrt{k}\l_s/(2\pi)$,
%and the coordinate $\xi dr/r = d \tau$.
%
While this is the   motivating example we
often have in mind, the strong
coupling singularity can lead to difficulties
and ambiguities in our conclusions.
We therefore wish to consider other examples
where the string coupling is
bounded above, although it can become zero as we approach the
boundary of the spacetime.

\bgn{\it Examples}

\item{1.} The doubly Wick rotated  near-extremal NS5-brane
\refs{\hs,\gibbons} \foot{ This background has a  tachyonic mode
due to the negative specific heat of the Euclidean black hole.
It will be clear that this is not important for our discussion.}
\eqn\nearext{\eqalign{
ds^2_{string} =& (1-r_0^2/r^2) d \phi^2 + (1+ { Q_5 \over r^2} )
\biggl[ {dr^2\over 1-r_0^2/r^2} +
r^2 ds^2_{S^3}\biggr] + (-dt^2  + ds^2_4)
\cr
e^{2 \phi -2\phi_{\infty} } = & 1 + { Q_5 \over r^2}
\cr
Q_5 =& r_0^2 \sinh^2  \beta ~,~~~~~~~ \alpha' k = r_0^2
{ \sinh 2 \beta \over 2}
}}

\item{2.}  The solution describing $k$  $NS5$ branes
wrapped on $S^2\times R^{3,1}$ \mn .
In this solution the structure of the spacetime near the boundary
is the radial direction times  $R^{1,3}  \times X_5$, where $X_5$ is
a (topologically trivial, but metrically nontrivial)
 $S^3 $ fibration over $S^2$. The full geometry is topologically
$R^{1,3} \times S^3 \times B^3$. In other words, the $S^2$ is filled
by a three ball in the full geometry. The radius of $S^3$ is constant and
there is a constant $H$ flux equal to $k$ over the three sphere.
Even though in \mn\ the type IIB version was mainly considered, we
can similarly consider the type IIA version of this background.
This is a supersymmetric background preserving four supercharges.

\item{3.} Geometries of the form $AdS_3 \times S^3 \times M_4$, where
$M_4$ can be $K3$, $T^4$ or $S^3 \times S^1$ with NS $H$ fields on
$AdS_3$ and $S^3$.

  Unfortunately, it is not obvious how to incorporate the
interesting solutions of Klebanov and Strassler \KlebanovHB\ since
they have RR fields and we do not know what replaces K-theory in
the presence of RR fields.

In order to measure the RR charge present in these backgrounds we
should   measure the flux of the RR fields at infinity, i.e. at
the boundary $\p X_9$. As argued in
\refs{\selfduality,\freedhopkins,\dmw,\freed} RR fieldstrengths
are topologically classified by K-theory, so the fields at the
boundary are classified by \eqn\fluxes{\eqalign{ \CQ_{IIA}^{RR} &
= K^0_H(\p X_9) \cr \CQ_{IIB}^{RR} & = K^1_H(\p X_9) }} where
$\CQ^{RR}$ are the charges  defined by measuring the RR fluxes at
the boundary. In the geometries we have considered the IIA fluxes
at infinity are
\eqn\twoachgp{ K^0_H(X_8) = K^1(X_5) \otimes
K^1_H(SU(2))\cong K^1(X_5)\otimes Z_k }
and are  therefore $k$-torsion.
In \twoachgp\ we used that  $K^0_H(SU(2))=0, K^1_H(SU(2))=Z_k$.

Let us now focus on the group of D0 charges as measured
by fluxes at infinity. According to
\twoachgp\ this group is
$Z_k$.   In this paper we show that this $Z_k$ conservation law can be
understood as a the result of a Higgs mechanism. A field with
charge $k$ gets a vacuum expectation value and therefore we only have
a $Z_k$ conservation law.
We will also show that this answer can be misleading in some cases.
In those cases there are ``singleton'' modes that can keep track
of $k$ units of charge and therefore the full charge group measured
at infinity is really $Z$. Whether this happens or not depends on
the behaviour of the metric and dilaton near the boundary and it
is not captured by K-theory which depends only on the topology of the
boundary.

It is important to bear in mind that in this
paper we are defining the charge group as
the possible fluxes that we can measure at infinity. This should
be distinguished from the topological classification of D-brane
sources (which is sometimes called the group of D-brane charges).
This latter group is given by the sets of topologically distinct
(in string field theory sense) D-brane configurations.
This is
given by the K-theory of the whole space (as opposed to the K-theory
of the boundary, which classifies the RR fluxes). In fact one can
connect the two by saying that \selfduality\
\eqn\chargegroup{
\eqalign{
\CQ_{IIA}^D & = K^0_H(\p X_9)/j(K^0_H(X_9))\cr
\CQ_{IIB}^D & = K^1_H(\p X_9)/j(K^1_H(X_9))\cr}
}
where $j$ is the inclusion map. The essential idea is
that this is a form of Gauss' law: Charges are measured by
fluxes at infinity. Thus the $K^1_H(X_9)$ charges usually
associated with IIA theory are measured by $K^0_H(\p X_9)$.
In accounting for the charges of D-branes we should
mod out by those classes which extend smoothly inside
since such classes
can exist without a D-brane source.
%
%, since
%a class which extends smoothly inside cannot arise from a
%D-brane in the interior.
%

\newsec{IIA spacetime interpretation}

In this section we explain how one can understand
the $Z_k$ D0 charge group in terms of IIA supergravity.
We then explain how the restoration of the $U(1)$
symmetry fits in.

\subsec{Higgs mechanism}

We will consider spacetimes $X_{10}$ which are fibered over
7-dimensional spacetimes $X_7$ by $S^3$, where  $S^3$ carries $k$
units of H-flux.  We will perform a Kaluza-Klein reduction along
the $S^3$ and keep only the lowest energy modes. In this section
we only keep the most important terms for the physical discussion.
Thus, the RR 3-form potential is reduced by
\eqn\kkredi{ C_3  =  \chi(x) \omega_3 +  c_3 }
 where $\chi(x)$ is
a scalar in $X_7$ and $c_3$ is pulled back {}from $X_7$.
Similarly, the RR 1-form $C_1\to c_1$ is assumed to be pulled
back from  $X_7$.

 As for the other fields in the theory, the $H$-flux
is frozen to be $H=k \xi_3$, and we neglect metric perturbations.
This ansatz can be justified in supersymmetric backgrounds
such as the throat geometry of the extremal fivebrane
since these have no tachyons. In other examples,
a more careful analysis is needed. Even if there are
tachyons we believe that our discussion captures the
main physics.

The RR sector of the IIA supergravity Lagrangian reduces to

\eqn\rrsector{
S= \int_{X_7} \vert d\chi + k c_1 \vert^2 + \vert dc_1\vert^2 +
\vert dc_3 \vert^2 + k c_3 dc_3.
}
The RR $U(1)$ gauge symmetry acts on $\chi$ because it acts on
$C_3$ in ten dimensions $\delta C = D_H \Lambda \equiv  (d -
H\wedge ) \Lambda$.  This gauge transformation law shows that
$\chi$ shifts in the appropriate way so that the order parameter
\eqn\chargedfield{ \Phi= e^{i \chi} = \exp\bigl[i \int_{S^3} C_3
 \bigr]}
has charge $k$.

{}From \rrsector\ it is clear that the $U(1)$ RR gauge symmetry is
spontaneously broken by the standard Higgs mechanism.
Furthermore, since it is broken by the charge $k$ order parameter
$\Phi=e^{i\chi}$, the $U(1)$ symmetry is spontaneously broken
down to $Z_k$. Since the dual group of $Z_k$ is again
$Z_k$, the $Z_k$ group of D0 charges is explained by a
completely standard physical mechanism.

\bigskip
{\bf Remarks:}

\item{1.}  There are  three distinct and
independent sources of mass terms in
the 7 dimensional theory which should not be confused with
one another.  First, the linear dilaton gives mass to some NS fields.
Second there is a gauge invariant mass term for the 3-form
$C_3$ in 7-dimensions giving this field a mass of order
$1/k$ see \rrsector . This mass term does {\it not} break the
$U(1)$ RR gauge symmetry associated with $c_1$. Neither  does
it break the gauge symmetry of the $c_3$ field.
Indeed, note that \rrsector\ is a sum of two decoupled systems.
Third, there is a {\it Higgs} type Lagrangian for the
coupling of $\chi$ to the RR vector field $c_1$. This is the term
responsible for the symmetry breaking and the consequent torsion
charge. In appendix A we will review several general facts about
Chern-Simons/BF theories, and in particular show that
the Higgs-type couplings are dual to the Chern-Simons-like terms.

\item{2.} Several authors have discussed several definitions of
charge \refs{\bds\PAH\marolf\figeroa-\taylor}.  In the viewpoint
advocated in this section, there is in fact no U(1) charge simply
because a charged field has obtained an expectation value, and we
cannot define a charge in a spontaneously broken vacuum.

\item{3.} In the usual discussion of the Higgs mechanism
a potential is responsible for fixing the modulus of the
Higgs field. It is
possible to make the modulus of $\Phi$ in \chargedfield\
dynamical and have a Lagrangian in which the $U(1)$ gauge
symmetry is realized linearly.  In such a Lagrangian a potential
must fix the expectation value of $|\Phi|$ to one and then this
massive field can be integrated out.  One would have to
perform a more detailed Kaluza-Klein analysis to learn about this
potential. For our purposes it is
enough to consider the simpler problem without a dynamical
$|\Phi|$ which is based on \rrsector.

\subsec{How to measure torsion charges}

One interesting question  is whether
 we can measure torsion charges at long distances. It seems clear
that if a charge is conserved in a quantum theory of gravity there
should be a way to measure it at long distances, otherwise the
charge can fall into a black hole and disappear.
In the present case we can measure these charges by measuring
Aharonov-Bohm phases in the following way.
For simplicity consider  the NS-5 brane, but the same could
be repeated for the other backgrounds mentioned above.
Suppose we want to measure the D0 brane $Z_k$ charge.
We consider a D4 brane that is a point on $S^3$. This
implies that in the extra seven dimensions the D4 brane has
codimension two. Then we take the D4 brane around the D0 branes
and we produce a phase
\eqn\phase{
e^{ i 2\pi  n/k}
}
where $n$ is defined as
the D0 brane charge. Clearly this charge is defined
modulo $k$ and furthermore, we can measure it at long distances.
This phase comes from the Chern Simons couplings of the RR gauge
potentials. More explicitly, in the ten dimensional
IIA supergravity Lagrangian there is a coupling of the form
$ \int  ~ d C_5 \wedge A \wedge H \sim k \int   ~d C_5 \wedge
A$, where $C_{5}$  dual to $C_{3}$. (In terms of $C_3$ this coupling
comes from the kinetic term which is
of the form $(dC_{3} + A\wedge H)^2 $).
 At long distances we can neglect the
kinetic terms for $C_{5}$ and $A$. In the presence of a D4 brane
the equation of motion for $C_5$ implies that the field  $A$
around the fourbrane is of  the form $A_{\varphi} \sim 1/k$ where
$\varphi$ is the angle around the fourbrane. So when we take the
D0 brane around the fourbrane we get the phase \phase . One can
similarly measure D2 brane charges by taking them around other D2
branes oriented in different directions. In principle it  should
be possible to compute this directly from the one loop open string
diagrams. This has indeed been done in \gukov\ for type I torsion
charges. It seems that one should be able to measure  NS torsion
charges via  similar  Aharonov-Bohm phases.
There are several interesting open problems related to these
questions.  For example, it would be nice to have a
simple K-theory formula for such phases.
\foot{A natural proposal for such an expression is the following.
Consider IIA theory on
$X_9 \times R_t $. Suppose a brane produces a
torsion flux. (It must be a torsion flux in order
to be able to speak of Aharonov-Bohm phases in the first place.)
Accordingly, we get an element of
$K^0_{tors}(X_8 )$ where $X_8 = \p X_9$. We are
going to measure the phase at infinity. Using the
exact coefficient sequence we lift the torsion element
to an element of $K^{-1}(X_8; U(1))$. Now, the
test brane defines an element of $K^0(X_8)$.
There is a natural pairing
$
K^0(X_8) \times K^{-1}(X_8;U(1))
\rightarrow K^{-1}(X_8;U(1))$.
Thus, to a torsion flux and test brane charge
we get an element of $K^{-1}(X_8;U(1))$.
Now let us consider the Aharonov-Bohm phase
for transport along a element of $H_1(X_8)$.
We lift this to an element of $K$-homology
and consider the pairing $
K_1(X) \times K^1(X;U(1))\rightarrow U(1)
$.
While this pairing exists on purely
topological grounds it can also be
defined in  terms of eta invariants of Dirac
operators
\lott, and these likewise enter in discussions
of Aharonov-Bohm phases. Thus, it is natural to
guess that this is the
Aharonov-Bohm phase. We thank E. Witten for a discussion
on this matter. }
%%%%

\newsec{M-theory picture}

All the type IIA backgrounds described in this paper have a
simple M-theory lift. We just need to add the M-theory circle
using the standard formulas for the uplifting.
Our conventions for the relation  of IIA theory and M theory
are the following. 11-dimensional spacetime is a
circle bundle with  a globally well-defined 1-form
\eqn\oneform{
\Theta = d\varphi_M  + C_{1,\mu} dx^\mu
}
We have $0\leq \varphi_M \leq 2\pi $ so that $\Theta$ is
normalized to $\int_{S^1} \Theta = 2\pi$ along the fiber.
Denote the fieldstrength by $R_2 = dC^{(1)}$.
The metric is taken to be
\eqn\metric{
ds^2_{(11)} =  e^{4\phi /3}\Theta^2 + e^{-2\phi/3} ds^2_{IIA}
}
with $\phi$ independent of the the fiber coordinate. Thus there is
a $U(1)$ isometry of the metric. We will  decompose the $G$-field as
follows:
\eqn\geefld{
G^{M-theory} = \pi^*(R_4) + \pi^*(H) \wedge \Theta
}
We have written the explicit pullback by $\pi: X_{11} \to X_{10}$ for
emphasis, but will henceforth drop it. This defines $R_4$.
Note that $d \Theta = \pi^*(R_2)$ is a basic form so the
last term is just part of the definition of $R_4$.
{}From \geefld\ we get the Bianchi identities:
\eqn\bianchi{
\eqalign{
dR_4 - H \wedge R_2 & = 0 \cr
d H & = 0 \cr}
}
If the line bundle
is trivial we have a globally well-defined 1-form $C_1$. Then
we can write
\eqn\solone{
R_4 = d C_3 - H \wedge C_1
}

It follows from \geefld\ that in the backgrounds under
consideration we  have $k$ units of flux of $G_4$ on $S^1 \times
S^3$. The circle size is bounded above and the geometry is
non-singular. {}From this perspective the U(1) D0  symmetry
 becomes translation along the 11th direction. The metric
and the field strength are translation invariant, so it might
come as a surprise that translation symmetry is broken. To
understand this phenomenon consider an M2 brane whose worldvolume
wraps the $S^3$.  Its phase is  given by
 \eqn\phase{
 e^{i \int_{S^3} C_3  } = e^{ i k (\varphi_M-\varphi_M^0) }
} where $\varphi_{M}^0$ is an arbitrary constant \foot{One should be more
accurate here. The $M$-theory 3-form can be viewed as a
Cheeger-Simons 3-character. It is a group homomorphism from the
group of 3-cycles in 11 dimensions to $U(1)$ such that, if
$\Sigma_3 = \p B_4$ then $\exp[i \int_{\Sigma_3} C_3 ] = \exp[i
\int_{B_4} G_4]$, where $G_4$ is a closed form with $(2\pi)$-integral
periods. The formula \phase\  is really the ratio of
characters for the 3-cycles $S^3 \times P$ and $S^3 \times
P_0$, where $P,P_0$ are two points on the M-theory
circle. In the extremal 5-brane one
could also fill in the 3-sphere in which case one
would find $e^{i k \varphi_M}$ up to exponential corrections in the
radius from the core of the 5-brane. }.
 The fact that we
cannot define the phase in a U(1) invariant fashion breaks the
symmetry from $U(1) \to Z_k$. This effect is very familiar in the
context of a two-torus
 with a magnetic
field where the translation group is similarly broken.\foot{ One
might ask whether in a case where we have flux on $S^4$ the
symmetry is broken. In this case it is possible to define the
phase  in an $SO(5)$ invariant fashion by choosing a four-
manifold inside $S^4$ whose boundary is the $M2$ brane, as it is
done in the familiar case of the $SU(2)$ WZW model, which has the full
symmetry of $S^3$. }
%We can think of these $M2$ branes as instantons which break the
%$U(1)$ symmetry to $Z_k$.
The fact that such M2 brane instantons break the $U(1)$ symmetry
is  intimately related to the
picture advocated in \mmsii .

Since we have a background with $G_4$ flux we would be tempted to
interpret it in terms of  M5 branes. Notice however that the
background is non-singular, so it is not obvious where the branes
are located. Nevertheless we can think of these backgrounds as
being the dual gravity description of a system of M5 branes with
a small transverse circle\foot{ The condition of the circle being
small is really the condition that the decoupling limit (or
nearly decoupling limit) is that of the IIA NS fivebrane, which
has an M-theory lift in the IR.} $S^1_{M}$. The $U(1)$ symmetry
associated with D0 charge corresponds to translations along the
circle $S^1_{M}$. Since the fivebranes are transverse to this
circle, it is clear that we will break the $U(1)$ symmetry. What
is a bit surprising is that a $Z_k$ seems preserved, as would be
the case if the branes were equally spaced.  In general we can say
that there is no evidence that the $Z_k$ is broken from the
gravity perspective, suggesting that the branes are equally
spaced. Furthermore, in the gravity backgrounds we are considering,
 there are no zero modes corresponding to the motion
of individual branes.

In example 2 above we can go even further and make an argument
showing that the branes are  dynamically restricted to be
equally spaced. The argument is as follows. Example 2 can be
thought of as arising from IIA NS5 branes wrapped on the $S^2$ of
the small resolution of the conifold giving rise to an effective
$d=4$ $N=1$ theory in the IR. When the radius of the 11th dimension is
large this theory seems to have moduli corresponding to the
position of the NS fivebrane on $x_{M}$ (which together with the
expectation value of the dynamical two form field $b_2$ on the
5-branes on $S^2$ gives a chiral multiplet). It was shown in
\superpotential\ that M2 branes stretched between the fivebranes
give rise to a superpotential. For large radius of $S^1_{M}$,
$R_{M}$, this superpotential has a minimum when the branes are
equally spaced. We conclude therefore that the theory has a
massive vacuum with a $Z_k$ symmetry. By holomorphy we expect
that this will be the case also for small $R_{M}$ when the dual
gravity description is valid. In summary, for example 2 the $Z_k$
symmetry should be an exact symmetry. In example 1 we do not know
of an argument that will tell us that the $Z_k$ symmetry survives
all possible non-perturbative effects. In fact example 1 is already
perturbatively unstable.

Finally let us discuss the case of the usual extremal fivebranes
that preserve 16 supercharges. In this case we can move all the
fivebranes independently and there is no reason for expecting a
$Z_k$ symmetry. Indeed, in the IIA gravity solution that
describes this system there is a strong coupling region. D0
branes can fall into this region and disappear. If the branes are
equally spaced, then we expect a  $Z_k$ symmetry in the full
non-perturbative theory. The precise symmetry group depends on
what the theory is doing at the IIA strong coupling singularity.

\newsec{D-brane probes}

D-brane probes are very useful in assessing the symmetries of a
background. Gauge symmetries of the background become global
symmetries of the probe theory. So an  interesting perspective on
the symmetry breaking $U(1) \to Z_k$ is provided by considering
a D2-brane probe in the 5-brane background. To be definite,
consider a probe in examples 1 or 2  whose  worldvolume
is along the directions where the IIA metric
is trivial.  So the probe will be a point on $S^3$.

The probe theory is a 3-dimensional $U(1)$ gauge theory. As is
standard, we dualize the photon by
 \eqn\dualiz{ F= dA  = * d \varphi_M }
We can interpret the scalar field $ \varphi_M$ as the
position of the M2 brane on the 11th circle \townsend, and
identify our global $U(1)$ symmetry as dual to the gauge $U(1)$.
Since the gauge group is $U(1)$, and not $\IR$, there are
instantons for the $U(1)$ gauge field, $dF \sim \sum_i
\delta^3(\sigma - P_i)$, which break the global $U(1)$ symmetry
\polyakov.

The duality \dualiz\ is valid in the IR for the probe field
theory regardless of whether the radius of the
$M$-theory circle is large or not, i.e.
regardless of whether we can use the eleven dimensional
description for the whole background.
 The instantons which correct the 3 dimensional theory are
obtained by considering the full theory. They consist of D2 brane
world volumes wrapping $S^3$. The $H$ field on $S^3$ gives them
magnetic charge under $F$. In the M-theory description, they are
$M2$ brane world volumes wrapping $S^3$ which have an amplitude
\eqn\instamp{
\sim \exp\bigl[- T_{M2} \vol(S^3) + i \int_{S^3} C_3^M\bigr]
}
where $C_3^M$ is the $M$-theory RR 3-form potential.\foot{
Note that in our examples
 the amplitude of \instamp\ is small only if $k/g_s$ is
large,  i.e. only if we are in the perturbative
IIA regime. }
Since
$
G_4 = dC_3^M = k \omega_3 \wedge d \varphi_M $, the phase of the
instanton amplitude \instamp\ becomes
$
e^{ i k(\varphi_M - \varphi_M^0) }
$.
Consequently, the instantons generate terms in the low energy
effective action which are proportional to
 \eqn\bpleeaa{ e^{-T_{M2} \vol(S^3) } e^{ik(\varphi_M - \varphi_M^0)}}
demonstrating an explicit breaking $U(1) \to Z_k$.

\bigskip
\ndt {\bf Remark}: We have focused on the instantons relevant for
writing the low energy effective action on the brane.
Nevertheless, it  is interesting to ask about the effects of
instantons which simultaneously wrap both the fiducial worlvolume
of the probe {\it together with} the $S^3$. Such instantons have
zero scale size because of a certain  ``bubbling phenomenon''
present for the harmonic map problem in dimensions larger than
two. Briefly, suppose we have a map $ F: M \to M \times M $ where
$M$ is an   $n$-dimensional manifold equipped with metric $g$ and
$B$-field. Consider the action
\eqn\iofeff{
I[F] := \int_M  \vol(F^*(E \oplus E))
}
where $E=g+B$. If $F$ has degree $(1,1)$ then for $n>2$ there is a
bubbling instability. Indeed, let $h(x)$ be a test function
mapping the ball $\parallel x^\mu\parallel \leq 1$ (in some
metric, in some coordinate patch) onto $M$ with degree 1.
Moreover, suppose that on the boundary $h$ maps to a single point
$P_0$. We can then extend $h$ to the rest of $M$ simply by
letting everything outside the ball $\parallel x^\mu\parallel
\leq 1$ map to $P_0$. Now we define
$h_\lambda(x):=h(x^\mu/\lambda)$ for $\parallel x \parallel \leq
\lambda$ and $h_\lambda$ maps everything outside this ball to
$P_0$. It is not difficult to see that the configuration is
unstable to shrinking $\lambda \to 0$ when $n>2$. This leads to
the bubbled configuration where $F(x) = (x,f(x))$ with $f(x)$
wrapping once around $M$ at $x=0$ and $f(x)=$ constant for
$x\not=0$.

\newsec{Restoration of the $U(1)$}

Let us return to the general backgrounds of section 1.1. We have
explained how K-theory predicts that the D0 charge group
in such backgrounds
is $Z_k$. Equivalently, the $U(1)$ gauge symmetry of the RR 1-form
of IIA theory is broken to $Z_k$. We have explained this in
terms of the Higgs mechanism, brane probes, and dynamically
generated symmetries in M-theory.  We will now explain that
these arguments can be  wrong.   More charitably, we will show that
they present a limited point of view.

More precisely, we will explain that, in certain circumstances, the
unbroken group should really be regarded as $U(1)$, and not
$Z_k$, and consequently the group of D0 charges is $Z$, and not
$Z_k$. The apparent contradiction with the K-theoretic
formulation is resolved when one understands that the ``symmetry
restoration'' to $U(1)$ involves degrees of freedom ignored in
the K-theory analysis.

%
%We will take the worldvolume of the brane to be compact. The
%spacetime far away from the brane is flat $R^{1,4} \times T^5$.
%We therefore expect the $U(1)$ D0 brane charge to be conserved.
%We expect that the group of fluxes is really $Z$ and not $Z_k$.
%We will show in this section that there is a RR collective
%coordinate that can carry $k$ units of D0 charge.
%

\subsec{RR collective coordinates}

One viewpoint on the symmetry restoration is that
there is a collective coordinate of the RR fields,
which arises since ``large RR gauge transformations''
which do not vanish at infinity should be considered
as global symmetries, and not gauge symmetries.

Recall that  $H=k \xi_3$ for some 3-form
$\xi_3$ with integral periods such that
$\xi_3 \to \omega_3 (1 + \CO(1/r)) $ at
infinity. The equations of motion imply that
$d\xi_3 =0$ and $d(*e^{-2\phi} \xi_3)=0$.
Moreover, we assume that  $\xi_3$ is   normalizable, which
implies that the fivebrane has finite volume.

Under these conditions there exists a normalizable collective
coordinate for the RR fields:
\eqn\collectivecoord{
\eqalign{
C_3  & =  \chi(t) \xi_3\cr
C_1 & = -{1\over k} \dot \chi(t) (1-e^{-2(\phi-\phi_{\infty})})dt  \cr }
}
Thus, $R_4 = -\dot \chi(t) e^{-2\phi} \xi_3 \wedge dt $
satisfies the Bianchi identity
$dR_4 - H\wedge R_2=0$.  The zeromode of $\chi(t)$
provides a solution of the RR equations of motion.
The normalizability of the zeromode of $\chi(t)$ is determined by
that of $\xi_3$.

Large global gauge transformations show that  $\chi(t)$ is a
periodic variable, with period $\chi \sim \chi + 2\pi$.
In the quantum mechanical theory, this periodic coordinate
has a  spectrum of discrete momenta, each unit of momentum carries
$k$ units of  D0 charge, which together with the previous
$Z_k$ form the D0 charge group which is $Z$. A more detailed
analysis of this point will be given in section 6.

Now we can see what was missing in the analysis of the previous
sections. First, in the discussion of the Higgs mechanism, in
section 2.1 the zeromode of the Goldstone boson for the RR 1-form
is normalizable and hence does not induce spontaneous symmetry
breaking $U(1) \rightarrow Z_k$. Second, in the   M2 probe
analysis of section 4,  the zeromode $\varphi^{M}_0$ in \bpleeaa ,
must be integrated
over, so the mode becomes dynamical and the
symmetry is unbroken. This mode is like an axion since it appears
in the instanton amplitudes \bpleeaa\ as a $\theta$ parameter.

In the M-theory description we are allowing a motion of the
M-fivebranes along the 11th dimension.
 In fact, the easiest way to  get
the zero modes \collectivecoord\ by uplifting  the IIA background
and considering an infinitesimal boost
 $\varphi_M \to \varphi_M + vt $, $t \to t + v \varphi_M$.
The
4-form transforms as $G_4 = k \xi_3 \wedge d\varphi_M + v k \xi_3 dt$.
The metric becomes
\eqn\boosti{\eqalign{ ds_{11}^2 \sim & e^{4\phi/3}(d\varphi_M +
vt)^2 + e^{-2\phi/3}\biggl[ - (dt + v d\varphi_M)^2+\cdots \biggr]
\cr \sim & e^{4\phi/3}\bigl[ d\varphi_M + v(1-e^{-2\phi})dt\bigr]^2 +
e^{-2\phi/3}\biggl[ - (dt)^2 +\cdots \biggr] }}
up to $\CO(v^2)$. This fixes $C_1$ and we can now
obtain $C_3$ from \geefld\ to obtain the
  new RR potentials. These  are just those
given in \collectivecoord\ above.
To the extent that we can think of this geometry as a
wrapped 5-brane, the 5-brane is wrapped on the 5-manifold
$X_5$ and propagates in time. Therefore, its mass is
finite iff $\vol(X_5)$ is finite at $r=\infty$.
If there are $k$ M5 branes transverse to the M-theory
circle with coordinate $\varphi_M$, and the branes have
a finite mass (e.g., because they wrap a finite
volume space) then these five-branes can absorb
momentum and move in the $\varphi_M$ direction.

What is a bit surprising from this point of view is that the
quantization of the singleton mode gives us $k$ units of momentum.
The reason is that the transformation $\chi \to \chi + 2 \pi$ in
the IIA variables corresponds to motion of all the branes by
$\Delta\varphi_M = 2\pi/k$ along the 11th circle.
 The fact that this is a symmetry is more evidence
that the fivebranes are equally spaced.
More precisely,
we should
consider
the wavefunction for the system including
the position of all 5-branes
$x_1, \dots, x_k$
around the M-theory circle.
We can separate the
center of mass degree of freedom
$x_{cm} = {1\over k} \sum_i x_i$ and accordingly
separate the wavefunction
\eqn\wvfunction{ \Psi(x_1, \dots, x_k) = e^{2\pi i q x_{cm}}
\Psi_q^{\rm rel}(x_i - x_j) }
where we have written a state in an eigenstate of total
momentum around the M-theory circle of momentum $q\in Z$.
The relative wavefunction only depends on $q\mod k$, through
the phases that appear when we change one of the
coordinates  $x_i \to x_i + 2 \pi $.
Therefore, when
 we change $q \to q+k$ there is no change in the relative
wavefunction and this mode is captured by a simple  collective
coordinate as above.
 On the other hand if we change $q$ by another amount
we need more information about the system to determine the relative
wavefunction.

\subsec{Some U-dual descriptions: KK-monopoles and H-monopoles}

An interesting perspective on the phenomena
discussed in this paper is provided by the
story of unwinding strings in the presence
of Kaluza-Klein monopoles  \unwinding.
This
example also illustrates the need for interpreting
\fluxes\ with care.

Consider first type IIB on  $R^{1,4}\times X_5$  with
flat metric on $R^{1,4}$ and zero $H$-field.  Then
$\p X_9 = S^3 \times X_5$.
The reader may set $X_5= T^5$ without much
loss of generality.  Let us consider the
RR fluxes  in such a background.
For IIB we should compute the flux group at infinity
\eqn\fluxgroup{ K^1(\p X_9) = K^1(S^3)\otimes K^0(X_5) ~ \oplus ~
K^0(S^3)\otimes K^1(X_5). }
We are  interested in fluxes associated to D1 branes
in $R^{1,4}$. These are measured (in cohomology)
by a degree seven class on the boundary
(dual to $H_3^{RR}$ in cohomology)
 which is of degree five  on $X_5$ and of degree two
on $S^3$. There is no such class in \fluxgroup . This is not
at all surprising since there are no one cycles in $R^4$ where the
D1 branes can be wrapped.

Now let us consider $R^{1,4}\times X_5$  but with a
KK monopole, or Taub-NUT  metric on $R^4$, and
with zero $H$-field.
Since the metric is smooth and since we can take the string coupling
to be arbitrarily
weak, the K-theory analysis is valid.
The topology of this space is the same as in the example above so
that  the RR fluxes are again \fluxgroup .
 In particular, if we view the sphere
$S^3$ at infinity as an $S^1$  Hopf fibration over $S^2$,
\fluxgroup\ predicts
there is no charge associated with the winding
number of D1 strings around the fiber coordinate.
Fluxes associated with such strings would be degree
7 classes associated with elements of
$K^0(S^3) \otimes K^1(X_5)$ which are degree two in $K^0(S^3)$,
 but there are no
such classes.

In fact,  there {\it is} a nontrivial
RR charge associated with the winding of D1 strings around
the Hopf fiber at infinity.
The apparent contradiction with K-theory is resolved
by noting that  the analysis leading to \fluxes  \fluxgroup\
neglects the
collective coordinate degrees of freedom in the RR
potentials explained in section 5.1.
The collective coordinate for the KK monopole and
the fact that it carries RR charge was discussed in
\unwinding \foot{
 In \unwinding\ fundamental strings were considered,
here we want to consider D-strings. The discussion is exactly the
same after exchanging $B_2^{NS} \to B_2^{RR}$ }.
The  collective coordinate of the KK monopole comes
from $B_2^{RR} = \alpha(t) \Omega_2$ where $\Omega_2$ is a
normalizable harmonic 2-form on TN \sencc .
This collective coordinate carries D1 string charge and the
charge measured at infinity is indeed $Z$, signaling
an unbroken $U(1)$ gauge group. Again, it is important that
$X_5$ is compact, otherwise this collective coordinate would
not be normalizable.
In conclusion, it seems that to determine the physically relevant
group of charges, as measured by RR fluxes at infinity, one needs
more information than just the topology of the space. As we
explained above, flat $R^{1,4}$ and the KK monopole have the same
topology, but we expect different physical charges at infinity.
It would be nice to understand how to correct \fluxes\ in a
general background, so
that it gives the physically appropriate answer.

As discussed in \unwinding\ it is also quite interesting to
consider the T-dual description of the above phenomenon. Now we
have type-IIA theory on an H-monopole. A wound D1 string becomes a
D0-brane, in the presence of an NS5 brane. The transverse space
is $R^3 \times S^1$. At long distances $H \sim \omega_2 \wedge d
\theta$ where $\omega_2$ is the unit volume form on the sphere
$S^2\subset R^3$. Therefore, the D0 can end on a D2-instanton
wrapping $S^2\times S^1$. Should we conclude that D0 brane charge,
as measured by the corresponding RR flux, will be undefined?
As we have seen in  section 5,
 the answer is ``no'': There is a
mode of the IIA RR potential $C_3 \sim \chi  \omega_2 \wedge d \theta $,
 for which our standard story applies: $\chi$ is the
Goldstone boson   eaten by the RR 1-form $C_1$. Using the explicit
form of the metric one checks that $\omega_2 \wedge d \theta$ is
normalizable in the weak coupling region of the
H-monopole.\foot{The above form is   non-normalizable in the
strong coupling region of an extremal $H$-monopole. In the
philosophy of this paper we should consider a smooth background
with this strong coupling singularity removed. Then the mode will
be normalizable.} This closes the circle of ideas.

In the above discussion we pointed out that the RR fluxes are
not properly classified by \fluxes . On the other hand we
have no complaints against  \chargegroup\  as a group of
D-brane sources.
In our case \chargegroup\ together with \fluxgroup\ gives us
the source charge group
$\CQ^D \cong K^1(S^3)\otimes K^0(X_5) \cong K^0(X_5)$.
This again vanishes for D1 branes
wrapping the fiber of the Hopf fibration, but this is not surprising
 since they  can
shrink to nothing. So
there is no problem in the interpretation of
\chargegroup\ as the group of topologically distinct sets of
D-branes that we can have in a given background.

The above discussion has been carried out for a singly charged
KK monopole. The generalization to charge $k$ monopoles
is rather interesting.  Let us therefore consider a smooth
Taub-NUT space $TN_k$ corresponding to $k$  KK monopoles.
The boundary at infinity is topologically a
 Lens  space $S^3/Z_k$, and metrically the
fiber of $S^3/Z_k \to S^2$ is asymptotically of constant
radius.  There
appears to be  a conserved $Z_k$ winding charge since two strings
winding $n$ and $n+Nk$ times around the fiber can be smoothly
deformed to one another at infinity. Moreover, for the resolved
TN space the fundamental group is $\pi_1(TN_k)=0$, suggesting
that there is no winding charge at all. Note that the unwinding
of $k$ strings can be done far away  from the cores of the KK
monopoles, but in order to unwind a smaller number of strings we
need to go to the cores of the KK monopoles.

Let us now consider the RR  charge in the K-theoretic
description.
The K-theory of 3-dimensional Lens spaces is easily
computed \atiyahbook:
\eqn\klens{
\eqalign{
K^0(S^3/Z_k) & = Z + Z_k\cr
K^1(S^3/Z_k) & = Z \cr}
}
The nontrivial classes in $K^0$
can be represented by the flat line
bundles associated to the unitary
representations of $\pi_1(S^3/Z_k) = Z_k$.
These line bundles extend to the full
space $TN_k$ as tautological line
bundles in the hyperkahler quotient
construction.
Thus, the group of RR fluxes
at infinity is
\eqn\rrflux{
(Z+Z_k)\otimes K^1(X_5)~ \oplus~
K^0(X_5)}
 The factor $Z_k \otimes K^1(X_5)$ is
the group of fluxes associated to strings winding
around the nontrivial elements of $\pi_1(S^3/Z_k)$.
So we would conclude from this that the RR fluxes
associated to D1 branes wrapping the fiber is
$Z_k$-valued.
 As we explained above the collective coordinate
degrees of freedom imply that this flux is actually
$Z$-valued. In fact, in this case there are many
normalizable harmonic two forms  $\Omega_2$,
which lead to  RR zero modes. There is essentially one for
each KK monopole.

It is interesting also to compute the group of
D-brane source charges by  modding
 out by the image of $j$ in
\chargegroup.
In this case the image of $j$ is nontrivial since, as we mentioned
above,  the $Z_k$ fluxes
can
be extended to the interior.
This implies that the group of source charges is again
 $K^0(X_5)$, so that
there
is no source-charge coming from D1 branes wrapping the fiber.
This is consistent with what we said above, since a D1 brane
wrapping the fiber can
shrink to nothing.

The T-dual of this charge $k$ situation brings us
back to the D0 charge group. The unwinding
of the D1 strings at infinity is
  T-dual  to the decay of $k$ D0 branes
due to a D2 brane instanton.
{}From the K-theoretic viewpoint we compute
(e.g. via the AHSS)
\eqn\klensi{
\eqalign{
K^0_H(S^2 \times S^1) & =  Z\cr
K^1_H(S^2 \times S^1  ) & = Z+ Z_k \cr}
}
with $H=k \omega_2 \wedge d \theta$. The
$Z_k$ summand in the second line corresponds
to D0 charge, and the $Z$ summand in the second
line corresponds to D2 branes wrapping $S^2$.
In this T-dual picture we see the $Z_k$ group in the weakly coupled
region, but the fact that it can be broken to nothing depends
on the behaviour of the theory in the strong coupling region, where
K-theory is not valid. However, by duality we know that in the
strong coupling region we should think in terms of $M$ fivebranes
and that the
$Z_k$ group is broken if we put the fivebranes at generic positions
in the 11 dimensional circle. It is also broken if we separate the
5-branes in the transverse three dimensional space. Separating
them in the three dimensional space is U-dual to resolving the
the singular charge $k$ KK monopole into the smooth Taub-NUT space
described above.

Finally, we note that one could combine the stories and
consider $k_2$ H-monopoles together with $k_1$ KK monopoles.
The relevant K-theory group is computed from
\eqn\klens{
\eqalign{
K^0_{H}(S^3/Z_{k_1}) & = Z_{k_1}\cr
K^1_{H}(S^3/Z_{k_1}) & = Z_{k_2}.\cr}
}
Here $k_1 k_2 = k \not= 0$ and
$H=k_2 x_3$ is normalized so that $\int_{S^3/Z_{k_1}} x_3 = 1$.
Note that \klens\ is nicely compatible with T-duality.

\newsec{A detailed analysis of the singleton mode and $U(1)$ symmetry
restoration}

In this section we focus on
the background given in example 1 and  we
compactify the five directions along the fivebrane. We
dimensionally reduce to the $r,t$ directions
and retain only the RR potentials $\chi$ and $C_1$ (and we
rename the latter potential to be $A$ in this section).
The resulting  Lagrangian is
 \eqn\lagrangian{ S
=  {1 \over 4 \pi  }\int dt \int_{r_0}^\infty  dr
   \Biggl[ \beta (r) \bigl( \dot \chi +   k A_0\bigr)^2
 -  \gamma(r) \bigl( \chi' +  k A_1\bigr)^2  +
  {1 \over  \alpha(r) } F_{0r}^2\Biggr]
 }
where
 \eqn\defalf{\eqalign{ {1 \over \alpha} = & { v_5 \over
(2\pi)^6 { \alpha'}^{3} } 2 \pi^2 r^3 (1 + {Q_5 \over r^2 }) (1 -
{ r_0^2 \over r^2 } ) \cr
\beta = &{ v_5 \over (2\pi)^2 {\alpha'} }
 { 1 \over  2 \pi^2 r^3 (1 + {Q_5 \over r^2 }) }
\cr \gamma = & { v_5 \over (2\pi)^2 { \alpha'} } {(1 - {
r_0^2 \over r^2 } )   \over 2 \pi^2 r^3 (1 + {Q_5 \over r^2 })^2 }
 }}
where $v_5$ is the volume of the directions along the fivebrane
at large $r$, and $(\alpha' k)^2= Q_5 (Q_5+r_0^2)$.

Before proceeding we do a duality transformation in $\chi$ to the
variable $\tilde \chi$ so that the lagrangian becomes
%
%%%%
%%%% i think there were some misprints here
%%%%
%%%%
 \eqn\lagradual{ S = {1 \over 2 \pi } \int dt
\int_{r_0}^\infty dr \Biggl[ { 1 \over 2\alpha } F_{r0}^2  +
{1 \over 2 \gamma} (\dot{\tilde \chi})^2 - {1\over 2 \beta} (
{\tilde \chi}')^2  +
 k \tilde \chi F_{0r}\Biggr] + { k \over 2 \pi } \int dt A_0
\tilde\chi (r=r_0) }
where $\tilde \chi$ is defined as $ C_5 = \tilde \chi \xi_{v_5} $
where $\xi_{v_5}$ is  fiveform along the volume of the fivebrane
normalized so that its integral is one.
Both $\chi$ and $\tilde \chi$ are periodic with periods $2\pi$.
$C_5$ is the field dual to $C_3$ in ten dimensions, $dC_5 \sim  *
(dC_3 + H A)$.
 The last term in \lagradual\ is a boundary term
necessary in order to have the appropriate boundary conditions at
the origin ($r=r_0$). More precisely, we want to have boundary
conditions at the origin such that $A_0$ is free, therefore we
need the boundary term to cancel a boundary contribution in the
variation of the action. We also impose that $\tilde \chi (r=r_0)
=$constant. When the time direction is a circle, large gauge
transformations determine this constant
\eqn\fixconst{
\tilde \chi (r=r_0) =2 \pi n/k .}
$n$ can be interpreted as the number of D0 branes at the origin.
For simplicity we will not put any D0 branes at the origin.
In this case the boundary condition is $\tilde \chi \in  2\pi Z$.
With these
boundary conditions the full ten dimensional solution is
non-singular at the origin.
%This comes from the fact that
%$\tilde \chi$ is related to a fiveform potential along the
%worldvolume of the brane. The $S^1$ in
%the  five-dimensional cycle is
%contractible in the full geometry, it contracts to a point at
%$r=r_0$.
The reason for this is that the
five-form  must  be smoothly extendible to the
entire geometry. In examples 1 and 3 of section 1.1 there is
a circle along the brane worldvolume which is contractible
to a point  at $r=r_0$. In example 2 the there is a
2-sphere along the worldvolume contractible to a point.

We add to the system a number of Wilson line observables which
are D0 worldlines of charges $q_i$  which we take for simplicity
to be purely temporal.  Then the $A_0$ equation of motion is
\eqn\eqnmota{
\partial_r({1 \over \alpha} F_{0r}) +
 k \partial_r \tilde \chi +
 2\pi  \sum q_i \delta(r-r_i) = 0
 }
where $q_i$ are integers.

We {\it define} the  D0 brane charge to be the  RR flux
measured at infinity:
 \eqn\charge{ Q \equiv { 1 \over 2 \pi \alpha} F_{r0}\bigl|_{r=+\infty} . }
By integrating \eqnmota\ we conclude that the total charge is
\eqn\chargetoto{ Q =k
 {(\tilde \chi(\infty) - \tilde \chi(0))\over 2 \pi }  + \sum_i q_i
 }

The value of $\tilde \chi(\infty)$ is a theta angle and it gives
a fractional value to the D0 brane charge of a fivebrane by the
usual Witten effect. Without much loss of generality we can set
it to zero, since it is just an overall shift in the charge. Note
that even though $\tilde \chi$ is only defined modulo $2\pi$, the
difference $ \tilde \chi(\infty) - \tilde \chi(0)$ is a
well-defined real number
 if $\tilde \chi$ is continuous. We then identify $k(
\tilde \chi(\infty) -  \tilde \chi(0))/(2\pi) $ as the
contribution to the charge from the singleton.

In order to gain a bit more insight notice that  we could think of
spacetime as divided into two regions, the region where the photon
is massive, which is the throat region and the region where it is
massless which is the asymptotic region far away from the brane.

In order to get some intuition on the behaviour of the system let
us understand the dynamics in the massive region. So let us set,
$\alpha, \beta, \gamma$ to constants $ \beta =1 $,
$\gamma=\gamma_0$ and $\alpha = \alpha_0$.
%%%%
%%%% i found the two distinct meanings of r somewhat confusing
%%%% and changed r to rho in the toy problem.
%%%%
%%%%
Moreover, we introduce a spatial variable
$\rho$, with $-\infty< \rho < \infty$.  Let us find the fields
produced by a Wilson line of charge $q$ in the time direction
inserted at $\rho=0$. The equation of motion for $\tilde \chi$, for a
time independent configuration, is
\eqn\motchi{
\partial_\rho^2 \tilde \chi  + k F_{0\rho} =0
} We can solve \eqnmota\ by setting \eqn\solva{ {1 \over
\alpha_0} F_{0\rho} + k \tilde \chi  + 2 \pi q \theta(\rho) =0 } The
solution is \eqn\solutions{\eqalign{ \tilde \chi = &2 \pi  {q
\over k} ( {1 \over 2} e^{- \kappa \rho } -1)~,~ ~~~~~ {1 \over
\alpha_0} F_{0\rho} =  - 2\pi q  {1 \over 2} e ^{ -\kappa \rho }
~~~~~~~~ \rho>0 \cr \tilde \chi = & - 2 \pi {q \over k}  {1 \over 2}
e^{- \kappa |\rho |} ~,~~~~~~~~~ {1 \over \alpha_0} F_{0\rho} = 2\pi q
{1 \over 2} e^{- \kappa |\rho| }  ~~~~~ \rho<0 }} where $\kappa = k
\sqrt{\alpha_0}$.

So we see that as we cross the Wilson line from $\rho<0$ to $\rho>0$
the value of $\tilde \chi$ jumps by $-2\pi q/k$. We can obtain the
long distance version of this result by considering the long
distance version of the lagrangian \lagradual\
 \eqn\long{ S \approx {1 \over 2 \pi } \int k \tilde \chi F }
which is a ``BF'' theory in two dimensions. (See appendix A for
some relevant facts about such theories.)

In conclusion, we find that even though the electric field
produced by a point charge decays exponentially, the $\tilde
\chi$ field ``remembers'' how much charge there was. Since
$\tilde \chi $ has period $2 \pi$  the charge is defined modulo
$k$ in the massive region.

In order to get some insight for the singleton, let us consider a
simplified system with a boundary at $\rho=0$. We need some boundary
conditions at $\rho=0$. We impose boundary conditions $A_0(\rho=0)=
\tilde \chi (\rho=0)=0$. The first boundary condition implies that
the $U(1)$ gauge transformation parameter is independent of time
at the boundary. The constant part is a global symmetry, i.e. we
only divide the path integral by gauge transformations where the
gauge parameter goes to zero at the boundary. The condition that
$\tilde \chi=0$ will be more fully justified later. Time
independent solutions of the resulting equations are of the form
\eqn\solsing{ \tilde \chi = \tilde \chi(-\infty) ( -e^{- \kappa
|\rho|} + 1) ~,~~~~ {1 \over \alpha_0} F_{0\rho} = k \tilde
\chi(-\infty)
 e^{ k \rho} ~~~~~~~\rho<0
} where $\tilde \chi(-\infty)$ is the value of $\tilde \chi$ as $
\rho \to -\infty$.
This is the singleton mode  that carries the
charge. The total charge is given by
\charge\ evaluated at the boundary, and this can
be read off from \solsing.
If there are no other charges in the interior, we should
use that $\tilde \chi(-\infty)/(2\pi) $ is an integer and then we
get that the singleton carries charge $k$. More precisely, in
situations where the fivebrane geometry is cut off at some finite
value of $r=r_0$ we argued above that it should be an integer.

%%%%
%%% i think some statemenst about gamma were wrong.
%%%%
This simplified model appears naturally in the theory of the
extremal fivebrane as follows. In that case,
  $\beta =\alpha/\alpha_0$ with $\alpha_0 = (2
\pi)^{8}\alpha'^4/v_5^2$. After defining a new variable $\xi$
through the equation $d \xi = \beta dr$ we find that we get a
theory like \lagradual\ with $\beta \to 1 $, $ \gamma \to
\gamma^2 $, $\alpha \to \alpha_0$. Explicitly,
\eqn\xivariable{
\xi - \xi_0 = {v_5\over (2\pi)^4\alpha' Q_5}
\log{r^2\over r^2+ Q_5}
}
and we choose $\xi_0$ so that the physical range of $\xi$ is
$ -\infty < \xi < 0$. The region of $\xi \sim 0$ corresponds to
the asymptotically flat region far from the fivebrane. The fact
that $\gamma$ is nonzero does not modify any of the time
independent equations we considered above. It does, however, have
an important consequence. First it implies that $\tilde \chi$ is
massless in the region $\xi \sim 0$. Second it implies that the
only reasonable
 boundary condition
we can impose on $\tilde \chi$ at $\xi =0$ is a Dirichlet boundary
condition. As we explained above this constant value of $\tilde
\chi(\xi=0)= \tilde \chi(r=\infty)$ is like a theta angle in the
full theory. Setting $\tilde \chi =0$ at $\xi =0$ we can compute
the energy contained in the singleton mode by inserting \solsing\
with $\tilde \chi(-\infty) = 2 \pi n $ into the Hamiltonian. We
find that the energy is
 \eqn\energy{ E = \half k n^2  2 \pi
\sqrt{\alpha_0 }   =
 \half k n^2 { M_0^2 \over M_5 }}
where $M_0$ and $M_5$ are the masses of one D0 and one NS5 brane.
This is, of course, the expected answer based on BPS bounds and
it can be thought of as the energy of $k$ M5 branes with $n k $
units of momentum in the 11th dimensional circle, in perfect
accord with the $M$-theoretic interpretation of the
singleton degree of freedom of section 5.1.

So- we can at last answer the question:
``What happens when $k$ D0 branes disappear?'' Since $\tilde \chi$ is
a periodic variable we have no physical effect if along some
trajectory $\tilde \chi$ shifts by $2 \pi$. Let us call such a
configuration a ``Dirac string.'' If we have $k$ coincident D0
branes, we also have this shift by $2\pi$ if we are far away from
the D0 trajectories. This suggests that $k$ D0-brane lines could
be replaced by a Dirac string, as in Fig. 1. In the low energy theory \long\
both are equivalent. In the original Lagrangian \lagradual, they
are not. In fact $k$ D0-brane lines cannot terminate due to
current conservation, which in turn
follows from gauge invariance. This
problem can be solved if we add a new object to the theory which
is the baryon vertex \baryons , where $k$ D0 lines can end. This object
acts as a magnetic source for $\tilde \chi$; it implies the
equation
 \eqn\magsource{ d d \tilde \chi  = 2 \pi  \delta^2(x)}
where $x=(r,t)$.
Note that the winding of $\tilde \chi$ around the point where it is
inserted is precisely the periodicity of $\tilde \chi$. Equation
\magsource\ also makes sure that the current conservation
condition is microscopically obeyed (so that we preserve gauge
invariance).  The simplest way to understand this is to go back
to the original variable $\chi$. If $k$ Wilson lines are ending
at the point $x=0$ there is a term of the form $e^{i k
\epsilon(x=0)}$ when we perform a gauge transformation
%%%
%%%% some normalizations were wrong below.
%%%%
$C_1 \to C_1 +
d \epsilon$. This is cancelled if the baryon vertex couples to
$\chi $ as
 \eqn\barcoup{ e^{i \chi (x=0)} }
so as to make the whole configuration gauge invariant.
(Recall that the covariant derivative is $d\chi + k c_1$. ) Then the
equation of motion of $\chi$ will be of the form $ \nabla^2 \chi
= 2 \pi \delta^2(x)$ which maps under the duality to
\magsource.\foot{The Laplacian is not that of flat space
but involves the functions $\beta, \gamma$ when these are
not 1. }
We can also see from the full string theory that the baryon
vertex couples to $\chi$ as in \barcoup. In this case the baryon
vertex is a D2 brane worldvolume wrapping the $S^3$. Since $\chi$
is related to the three form on $S^3$ as in \kkredi\ the coupling
of the three form to the D2 brane translates into \barcoup.

In conclusion, the singleton mode is responsible for the
conservation of the U(1) charge.

\ifig\baryon{ Here we see $k$ D0 lines ending on a baryon vertex.
A ``Dirac string'' for the $\tilde \chi$ field, indicated by the
dotted line, emanates from the baryon vertex. }
{\epsfxsize1.5in\epsfbox{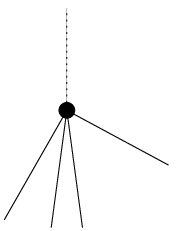}}

\bigskip
\noindent
{\bf Remarks}:

\item{1.} It is crucial,
for the $U(1)$ to be restored, that the worldvolume
of the fivebrane is finite, otherwise the 3-form would not
be normalizable.

\item{2.}
The presence of the singleton mode is also related to the fact
that Chern Simons theories  have physical propagating modes on the
boundary when spacetime has a boundary and we impose local
boundary conditions for the gauge fields. This is the relation
to the general theory outlined in appendix A.

\newsec{U-Duality in $AdS_3 \times S^3 \times T^4$}

First let us start with a flat space compactification  of IIB
string theory on $R^6 \times T^4$. The charges of particle like
excitations in six dimensions form a lattice $\CP \cong Z^{16}$.
There are 8 NS charges and 8 RR charges. Similarly there is a
lattice of string like charges in six dimensions $\CS \cong
Z^{10}\cong II^{5,5}$ . As representations of $SO(5,5;Z)$ the
particles are spinors and the strings are a vector. When we have
a  BPS string carrying some charge we can take its near horizon
limit, which will generically be $AdS_3 \times S^3 \times T^4$.
This system has been much studied. A partial list of relevant
references includes
\refs{\MaldacenaRE\MaldacenaBW\DijkgraafGF\MaldacenaBP
\KutasovXU\SeibergXZ-\larsenmartinec}.

A choice of near-horizon limit is a choice of string charge $\CS$.
We regard the ${\bf 10}$ as a symmetric bispinor
\eqn\grouptheory{ ({\bf 16} \otimes {\bf 16})_{\rm symm} \supset
{\bf 10} \oplus \cdots }
and denote it by $\CS_{\alpha\beta} = \CS_{\beta\alpha}$.
The gamma matrices acting on the spinor representation
$Z^{16}$ are integral, so \grouptheory\ can holds for
representations of $SO(5,5;Z)$.

On $AdS_3$ we will have 16 vector fields corresponding to the
sixteen vector fields that we had in six dimensions. These vector
fields have a Chern-Simons coupling given by
\eqn\gencsact{
S = \CS_{\alpha\beta} \int A^\alpha d A^\beta
}

The gauge group for the Chern-Simons fields is $U(1)^{16}$.
Singular  gauge transformations such as those
described in appendix A induce nontrivial Wilson loops.
The group of charges for Wilson loops may be elegantly
expressed as follows.
 From \grouptheory\ we see that
a choice of string charge $Q_{\alpha\beta}$ determines a map
$\Lambda_Q : \CP \to (\CP)^*$ by $p^\alpha \to  g^{\alpha\beta}
Q_{\beta \gamma} p^\gamma $ (where $g_{\alpha\beta}$ is the
lattice metric). The group of charges is
\eqn\wilsonchrg{
(\CP)/ \Lambda_Q(\CP)
}
This is  the finite group $(Z_{\half \CS^2})^8$ and carries a
signature $(4,4)$ quadratic form. In general there is no
invariant meaning to the level of the WZW model, and one obtains
different levels by going to different asymptotic regions. The
only invariant combination is $\CS^2$. One particular asymptotic
region makes the nature of the group particularly transparent.
Consider $Q_5$ NS5 branes and $Q_1$ parallel fundamental strings.
Thus, in some basis the string charge is $\CS = (Q_1, Q_5, 0,
\dots)$. We assume this is a primitive vector, or there are other
complications. Thus $Q_1,Q_5$ are relatively prime. Then we can
interpret the charge group as follows: \foot{We thank E. Witten
for a useful discussion on this subject.}

\item{1.}
The 4 D1 charges are broken to $Z_{Q_5}$ by Euclidean D3's
wrapping $S^3 \times T^1$.

\item{2.}
The 4 D3 charges are broken to $Z_{Q_5}$ by Euclidean D5's
wrapping $S^3 \times T^3$.

\item{3.}
The 4 F1 charges are broken to $Z_{Q_1}$ by KK monopoles  with
$T^1\subset T^4$ as the Hopf circle.

\item{4.}
The 4 momentum charges are broken to $Z_{Q_1}$ by NS5-branes
wrapping $S^3 \times T^3$, with $T^3$ orthogonal to the circle
carrying the momentum.

The above  gives a presentation of the finite group \wilsonchrg\ as
$(Z_{Q_1})^8\oplus (Z_{Q_5})^8\cong (Z_{Q_1 Q_5})^8$.
Nevertheless, the  true charge group is still $Z^{16}$, by the considerations above.
Note that the finite group $(Z_{Q_5})^8$ is the one deduced from $K_H(\p X_9)$.
This example should provide a useful test case for understanding the
proper formulation of the classification of fluxes together with
charges.

\newsec{Discussion}

In some discussions of charges of D-branes the issue of torsion
charges takes on an aura of mystery, especially in the K-theoretic
context. We would like to emphasize that there is nothing terribly
mysterious or exotic about torsion charges. They are present in
ordinary laboratory physics in, for example,  superconductivity.
We have seen that the mechanism by which such torsion charges arise
in 5-brane backgrounds is
simply through Higgs condensation
of a charged field which does not carry the fundamental unit of charge.
%  Unfortunately, it appears that
%not all torsion D-brane charges can be produced in
%this simple way, a counterexample being the type I nonbps D0-brane.
We can measure these torsion charges by measuring Aharonov-Bohm
phases at infinity.
%%%%
%%%%
%It would be nice to see if there is a nice
%expression in K-theory for these phases.

We further pointed out that the physical charge measured at infinity
seems to be different from the expression naively deduced from the
K-theory of the boundary of the space. The reason is that the
definition of the charge far away depends on the {\it metric} and
not just on the topology of the space. This is clearly illustrated
by the KK monopole example.
RR charges as measured by RR fields at infinity
can be carried by RR zero modes as well as by D-branes.
We believe this will be an important point in answering the
open problem mentioned in the introduction.

Many interesting issues remain open to future investigation.
It would be nice to have a clear and systematic criterion
for deciding when the $U(1)$ symmetries are, or are not,
restored. In example 2 of section 1.1 we believe the $U(1)$
symmetry is {\it not} restored, since the relevant
harmonic three form turns out to be non-normalizable. The D0 charge group
is then  $Z_k$ and not $Z$.

The issues we have discussed are intimately related to
the topological classification of
branes together with fluxes, a
subject which has not yet been adequately understood.
Finally, the issues we have discussed are related to
the relation between
$U$-duality and $K$-theory. The
relation of K-theory to D-brane instantons
described in \mmsii\ suggests a formulation of
an $SL(2,Z)$ invariant version of the AHSS which
is relevant to that problem, but we will leave a
detailed description of this to another occasion.

\bigskip
\centerline{\bf Acknowledgements}

We would like to thank E. Diaconescu, M. Douglas,  and
E. Witten for many conversations on
topics related to this paper over the past few years.

GM and JM  would like to thank the ITP at Santa Barbara for
hospitality during the writing of part of this manuscript.
This research was supported in part by the National Science
Foundation under Grant No. PHY99-07949 to the ITP.

GM is supported
by  DOE grant \#DE-FG02-96ER40949 to Rutgers. He also thanks
the Aspen Center for Physics for hospitality during the
conclusion of this work.
%
%The ACP is supported by
%NSF grant ??????
%

JM and NS were supported in part by DOE grant \#DE-FG02-90ER40542,
JM was also supported in part by DOE grant \#DE-FGO2-91ER40654,
NSF grants PHY-9513835 and the David and Lucile Packard
Foundation.

\appendix{A}{Some comments on Higher-Dimensional Chern Simons/BF
theories, and their role in supergravity  }

At several points in this paper we have relied on Chern-Simons and
BF type theories, which are generalizations of the basic
paradigm of three-dimensional Chern-Simons theory
\wittenjones. Here we collect a few relevant remarks on these
theories useful for following the main text. Much of what we say
is well-known to those who know it well, and can be found in an
extensive literature.

\subsec{Duality of  Chern Simons and Higgs Lagrangians}

In equation \rrsector\ we encountered the two types of systems
we would like to study. These Lagrangians are easily generalized,
and we will study the more general systems.

Let us begin with the general Higgs-type Lagrangion for $p$-form
gauge potentials in D-dimensions. The Minkowskian action is
\eqn\lagrahiggs{ \exp\Biggl[-i  \int_{X_D} \half \kappa_1 ( d
\lambda_p + k A_{p+1}) \wedge * ( d \lambda_p + kA_{p+1})
 +\half \kappa_2 d A_{p+1} \wedge * d A_{p+1} \Biggr] }
where $\kappa_1,\kappa_2$ are some positive constants. The
Euclidean theory has the same action without the overall factor
of $i$.

Consider first the local physics of \lagrahiggs.
The classical equations of motion
of the theory \lagrahiggs\ are:
\eqn\lagrheom{
\eqalign{
d\bigl[* (d\lambda_p + k A_{p+1})\bigr] & =0 \cr
d(*dA_{p+1}) + (-1)^p{k\kappa_1\over \kappa_2} *(d\lambda_p +
k A_{p+1}) & = 0 \cr}
}
The local gauge symmetry includes $A_{p+1} \to A_{p+1} + d\Lambda_p,
\lambda_p \to \lambda_p - k \Lambda_p$. After a gauge transformation
removing $\lambda_p$
we recognize the formulae of a massive field with mass-squared
$k^2 \kappa_1/\kappa_2$.  In Minkowski space,
wavefunctions are in the induced
representation from the antisymmetric tensor of rank
$(D-p-2)$ (equivalently, rank $p+1$) of $SO(D-1)$.

We will now dualize $\lambda_p$ to $\lambda_{D-p-2}$ to obtain
an equivalent formulation \lagrahiggs\ in terms of BF
theory, but before doing so we should discuss the global
nature of the fields.

The global nature of the fields in this theory is of the utmost
importance to its proper interpretation. \foot{At this point the
reader might wish to rotate to Euclidean space and take $X_D$ to
be compact.} Although we write gauge potentials $A_{p+1}$ and
$\lambda_p$, when $X_D$ is topologically nontrivial we wish to allow
field configurations where they are not globally well-defined.
Thus we allow the fieldstrengths $F_{p+2}=dA_{p+1}$ and
$f_{p+1}=d\lambda_p$ to be closed forms with nonzero periods.
These periods are $2\pi$ times an integer, and any integer can
appear. Thus the integral constant $k$ appearing in \lagrahiggs\
is \ meaningful. \foot{Some readers will therefore declare that
our fields should really be considered to be Cheeger-Simons
differential characters \cheegersimons. Useful expositions in the
physics literature can be found in
\refs{\gawedzki,\dijkgraafwitten,\hopkinssinger}. }

Global aspects are also crucial
 in the choice of the  gauge group. When $X_D$ is compact, the
choice most suitable to physics appears to be to allow
large gauge transformations $A_{p+1} \to A_{p+1} + \zeta_{p+1}$
where $\zeta_{p+1}$ is any closed form with $(2\pi)$-integral
periods. We will denote the space of
such forms by   $Z^{p+1}_{2\pi Z}(X_D)$.
Similarly, we take as gauge symmetry
$\lambda_p \sim \lambda_p + Z^{p}_{2\pi Z}(X_D)$.
It is for this reason that we cannot simply shift away
$\lambda_p$ by a gauge transformation in \lagrahiggs.

We are now ready to dualize $\lambda_p$ to $\lambda_{D-p-2}$. At
the classical level we can dualize  $\lambda_p$ into
$\lambda_{D-p-2}$ by interchanging equations of motion with
Bianchi identities. The equations of motion are solved for
\eqn\dualitytmn{\eqalign{
 * ( d \lambda_p +k A_{p+1}) & = -{1\over 2 \pi \kappa_1} d
 \lambda_{D-p-2}\cr
 * dA_{p+1} & = (-1)^p{k\over 2\pi\kappa_2} \lambda_{D-p-2} + \zeta_{D-p-2}
 \cr}
}
where $\zeta_{D-p-2}$ is a closed form (this is the constant $q$ of section 6).
%
%d \lambda_p + kA_{p+1}  = & * d \lambda_{D-p-2}
%}}
%
%% juan: the two equations you wrote are the same, just apply *
%
%

We perform the dualization quantum mechanically by following a
standard procedure: We introduce an auxiliary $(p+1)$-form
$B_{p+1}$ with no restriction that it be closed or have integral
periods, together with a new gauge potential $\lambda_{D-p-2}$
whose fieldstrength has $(2\pi)$-integral periods, and consider
the action
\eqn\lagrahiggsii{
\eqalign{
\exp\Biggl[-i  \int_{X_D} \biggl\{ \half \kappa_1 ( B_{p+1} + k A_{p+1})
& \wedge * ( B_{p+1} + kA_{p+1})
 +\half \kappa_2 d A_{p+1} \wedge * d A_{p+1} \biggr\} \cr
 & - {i\over 2\pi} \int_{X_D} B_{p+1} d \lambda_{D-p-2} \Biggr]\cr} }
Integrating out $\lambda_{D-p-2}$ forces $B_{p+1}$ to be closed
with $(2\pi)$-integral periods, and we recover the theory
\lagrahiggs\ together with its solitonic sectors. On the other
hand, we could also shift $B_{p+1}$ by $kA_{p+1}$ and perform the
Gaussian integral on $B_{p+1}$ to obtain the theory
\eqn\lagracs{
\eqalign{
\exp\Biggl[-i  \int_{X_D} \biggl\{ {1\over 8 \pi^2\kappa_1} d\lambda_{D-p-2}
& \wedge * d\lambda_{D-p-2}
 +\half \kappa_2 d A_{p+1} \wedge * d A_{p+1} \biggr\} \cr
 & - {i k \over 2\pi} \int_{X_D} A_{p+1} d \lambda_{D-p-2} \Biggr]\cr}
}
The new theory has a topological BF-type coupling.
In the Euclidean signature the first term is real but the BF coupling
remains imaginary.
In both theories \lagrahiggs\ and \lagracs\
 the number of degrees of freedom is the same, it
is that of a massive $A_{p+1}$ form. In this sense
the Higgs mechanism and Chern-Simons couplings are
dual to each other.

An important special case of the above theory occurs
in odd dimensions $D=2n+1$ with $p+1=n$. In this case
 we get a Chern-Simons selfcoupling and further discussion is
required.

In the present paper the most important examples are

\item{1.} $D=7, p=0$. The Higgs system is replaced by the
BF system with $\lambda_5, A_1$.

\item{2.} $D=7, p=2$. This is the self-dual 3-form.

\item{3.} $D=2, p=0$. This is the 1+1 dimensional theory
analyzed in detail in section 6.

\subsec{Topological theory in the long-distance limit}

The long-distance/low-energy dynamics of the  theory \lagrahiggs\
is  dominated by the BF coupling in the formulation
\lagracs. This is most easily seen by simply noting that
in \lagracs\ the BF coupling has only one derivative.
\foot{At this point some readers might get confused.
Usually, in justifying the dominance of topological
terms in an action one introduces a family
of metrics $g_{\mu\nu} = t^2 g_{\mu\nu}^{(0)}$  and takes
a  $t\to \infty$ limit. In our case
 we must also scale the potential $A_{p+1}$, or,
better, its gauge coupling $\kappa_2$. }
Thus we obtain a topological field theory of BF type:
\eqn\bftheory{
\exp\Biggl[ - {i k \over 2\pi} \int_{X_D} A_{p+1} d A_{D-p-2} \Biggr]
}
where we have renamed $\lambda_{D-p-2} \to A_{D-p-2}$ in this subsection.

Another example of this
phenomenon was studied in detail by Witten in \wittenadstft\ in the
context of the AdS/CFT correspondence on $AdS_5 \times S^5$.
In that case, the large $N$ limit jutified the condensation of
$\int_{S^5} G_5= N$ leading to a 5-dimensional TFT of BF type.
While the mathematics is rather similar to what we are discussing
in this paper, the physics is slightly different, since we obtain
a topological field theory by freezing a Neveu-Schwarz sector field.

The topological field theory \bftheory\ is more subtle than might
at first appear. Once again, it is absolutely crucial to specify
the global nature of the fields and their gauge symmetries. We
allow the on-shell fieldstrengths to be closed forms with
$(2\pi)$-integral periods. What should we take to be the gauge
group? Let $B^p$ be the space of exact $p$-forms and $Z^p$ the space
of closed $p$-forms. There are three natural choices of gauge group for the
field $A_i$:

\item{1.} $A_{i} \sim A_{i} + B^{i}(X_D)$.

\item{2.} $A_{i} \sim A_{i} + Z^{i}_{2\pi Z}(X_D)$.

\item{3.} $A_{i} \sim A_{i} + Z^{i}(X_D)$.

Accordingly, there are {\it six}   theories with Lagrangian
\bftheory. These theories are distinct. For example,
if both potentials are of type   1, $k$ is not quantized. If both are
of type  2, $k$ is quantized and integral. This is the choice
which appears to be most relevant to supergravity.
\foot{Thus, some readers will insist that the theory \bftheory\
is the gauge invariant product of a dual pair of
Cheeger-Simons characters. For a careful description of such
products see \harris.}
If both are of type 3, we must restrict the periods of
$A$ or of $\lambda$ to be zero.

The nature of the gauge group similarly restricts the possible
observables (``Wilson surfaces'') and the identifications imposed
between different observables by gauge symmetry. In theory 1, any
function of $\int_{\gamma} A_{i}$, where $\gamma$ are any closed
cycles, is gauge invariant. In theory 2, the observables are
\eqn\wilson{
W(\gamma) = \exp[i \int_{\gamma} A_i ]
}
where $\gamma$ is a cycle defining an {\it integral} homology class.
The cycle $\gamma$ generalizes the electric coupling of the gauge
field, and some readers will prefer to write $\gamma = e \gamma_0$ where
$e$ is an integer and $\gamma_0$ defines a primitive integral
homology class.
In theory 3, there are no gauge invariant observables.

The correlation functions of these observables are easily computed.
Let $\gamma_1, \dots, \gamma_n$ be integral $(p+1)$-cycles and
$\sigma_1, \dots, \sigma_m$ be integral $(D-p-2)$-cycles. Then
\eqn\correlators{
\biggl\langle W(\gamma_1) \cdots W(\gamma_n)\cdot W(\sigma_1)
\cdots W(\sigma_m) \biggr\rangle
= \exp\biggl[ {2\pi i \over k} \sum_{r,s} L(\gamma_r, \sigma_s)\biggr]
}
where $L(\gamma,\sigma)$ is the integral linking number.
\foot{The reader who is still awake might notice that we  have
neglected self-intersection terms. These require a choice of
framing of the normal bundle. In the situations of interest in
this paper the normal bundle will be trivial so we can take the
self-intersections to be zero. }
One way to interpret this formula is that the insertion of a
Wilson line along $e\gamma_0$ creates a holonomy
$2\pi e/k$ for the field $A_{D-p-2}$ around $\gamma_0$.

In theory 2, the Wilson surface observables are subject to an important
identification rule, namely
\eqn\idntify{
W(\gamma) \sim W(\gamma + k \gamma')
}
where  $\gamma'$ is any cycle
defining an integral homology class. This is plainly compatible with
the explicit correlators \correlators\ but it can also be derived
by the technique of making a ``singular gauge transformation,''
familiar from discussions of 3D Chern-Simons gauge theory, as well
as from the fractional quantum Hall system. In this case, one can
see that the effect of inserting
a Wilson line of charge $k$ is just performing
 a gauge transformation
$A \to A + 2\pi d \theta$ where $\theta$ is an angle around the
Wilson line to be inserted. These singular gauge transformations
are allowed in the theory, they are unobservable Dirac strings.

We now describe these singular gauge transformations in the
general case.
Let $\sigma $ be a closed integral cycle in $X_D$ of dimension $D-p-2$.
Let $\zeta_{p+1}$ be a trivialization of the poincare dual of
$\sigma $ on $X_D  - \sigma $, i.e. a solution of
$d\zeta_{p+1} = \eta(\sigma\hookrightarrow X_D)$ on
$X_D - \sigma$.  Thus, near
$X_D$,  $\zeta_{p+1}$ is a global angular form on the normal
bundle of $\sigma_{D-p-2}$. In 3D Chern Simons theory we have
$\zeta_{p+1} = d \theta$ where
$\theta$ is the angle around the Dirac string. If we have a
Wilson surface observable of the form
$ e^{ i k \int_{\sigma} A_{D-p-2} }$ we can see from the equations
of motion of \bftheory\  that the field configuration for
$A_{p+1}$ around it is the
same as that of a singular gauge transformation of the type we have
just described. Therefore the insertion of such a Wilson surface operator
is unobservable, since it is equivalent to performing a singular gauge
transformation, i.e. it is the same as adding an unobservable
``Dirac surface.''

%
%In 3D Chern-Simons theory with a
%Wilson line, $\zeta_{p+1}= d\theta$ where $\theta$ is the
%angle around the Wilson line.
%%%%
%%%%
%%%%
%The effect of the singular
%gauge transformation is to insert a Wilson surface observable of the form
%$\exp[i k \int_{\sigma  } A_{D-p-2} ] $.  We can see from the equations
%of motion of \bftheory\ that the field configuration produced
%around it is the same as that produced by a singular gauge
%transformation $A_{p+1} \to A_{p+1} + 2\pi \zeta_{p+1}$.
% We conclude therefore that the insertion of
% such operators is unobservable, that is, it is a
%``Dirac surface'' and hence we have the identification
% \idntify .
%%%%

\subsec{Theory on a manifold with boundary: The singleton degrees of
freedom }

Let us now introduce a boundary into the theories \lagrahiggs\ and
\bftheory.

We begin with the theory \lagrahiggs\ with local degrees
of freedom.
Let us also begin with a ``cylindrical spacetime'' by which
we mean a spacetime of the form  $X_D= \IR_t \times M_{D-1}$
where $M_{D-1}$ is a manifold with boundary $\p M_{D-1}= \Sigma_{D-2}$.
One should have in mind the example of the cylinder with $M_{D-1}$ the
ball of dimension $D-1$. Let $r$ be a normal coordinate near the
boundary so that ${\p \over \p r}$ is a unit normal vector.
A  system of boundary conditions such that
\lagrahiggs\ is a well-posed variational problem is
\eqn\bdryconds{ \eqalign{ \delta A_{p+1}\vert_{\p X_D} & = 0 \cr
\biggl\{\iota({\p \over \p r})[d\lambda_p + k
A_{p+1}]\biggr\}\vert_{\p X_D} & = 0 \cr} }
(the second line is simply the normal covariant derivative). The
boundary condition in the first line fixes the gauge symmetries.
We can no longer make gauge transformations by $\Lambda_p$ on the
the boundary, and consequently, in addition to the {\it massive} degrees
of freedom propagating in the bulk, there is a {\it massless} $p$-form
field propagating along the boundary. This is closely related to the
singleton degree of freedom. An analogous story holds in the
dual formulation \lagracs.

Let us now turn to the topological theory. In this case there are
two conceptually different choices for the manifold $X_D$ with
boundary. \foot{Cheeger-Simons characters for manifolds with
boundary have been studied in \zucchini. In the context of
theories of self-dual forms they have been studied by Witten in
\refs{\imwitten,\wittenadstft,\duality}. In the context of M-theory with
boundary they have been studied in detail in \dmunpub.}

First, if $\p X_D$ is compact then the path integral defines a
vector in a  Hilbert space. That Hilbert space is the quantization
of a finite-dimensional phase space. In general the phase space
is a quotient of $H^{p+1}(\p X_D;R) \oplus H^{D-p-2}(\p X_D;R)$.
When $D=2n+1$ and $p=n-1$ the phase space is a quotient of
$H^n(\p X_D;R)$. \foot{In the K-theoretic quantization of fluxes
this phase space is replaced by the K-theory torus, as studied in
\refs{\duality,\selfduality,\dmw}.} Different vectors in the phase space
are obtained by including different operators $W(\gamma)$ in the
interior of $X_D$. Nontrivial operators on the Hilbert space are
obtained from Wilson surface operators associated to cycles
$\gamma$ piercing the boundary $\p X_D$.

Second, we can consider \bftheory\ on a cylindrical spacetime
$X_D= \IR_t \times M_{D-1}$. In this case the theory is
holographically dual to a quantum field theory living on the
boundary. These dynamical degrees of freedom on the boundary are
sometimes referred to as ``singleton'' degrees of freedom,
because of their role in AdS theories. Now, the ``singular gauge
transformations'' act nontrivially on the Hilbert space of these
theories, and should not be considered to be gauge
transformations. Rather they produce modes with nontrivial
fluxes for the singleton degrees of freedom.
 The Hilbert space includes a sum over these flux
sectors. There are also new operators in the theory arising from
Wilson surface operators for surfaces which pierce the boundary.
These can be interpreted as insertions of charged operators
in the boundary theory.

One chooses boundary conditions so that variations satisfy
\eqn\bdrycond{ {k \over 2 \pi }
\int_{\IR_t \times \Sigma_{D-2}} A_{p+1} \delta
A_{D-p-2}=0. }
A convenient boundary condition which satisfies \bdrycond\ is
 \eqn\tempba{
\iota({\p\over \p t}) A_{p+1}=\iota({\p\over \p t})A_{D-p-2} =0}
where these are the components of the gauge fields which have an
overlap with the ``time direction'' $\IR_t$.  These boundary
conditions break some of the gauge symmetry leaving unbroken the
subgroup of gauge transformations which are zero on the boundary
$\Sigma_{D-2}\times \IR_t$.

In order to identify the spectrum on the boundary we follow a
procedure used in \refs{\wittenjones,\emssr}.  We note that the
time components of the gauge fields in \bftheory\ are Lagrange
multipliers, so integrating over these we impose the constraints
\eqn\constraints{ \eqalign{ d(A_{p+1}\vert_{M_{D-1}}) & = 0 \cr
d(A_{D-p-2}\vert_{M_{D-1}}) & = 0 \cr} }
If $H^{p+1}(M_{D-1})=0, H^{D-p-2}(M_{D-1})=0$, we can globally solve
$$A_{p+1}\vert_{M}= d^{(D-1)} \Phi_{p} , A_{D-p-2}\vert_{M}=
d^{(D-1)} \Phi_{D-p-3}$$ and substitute back into the
Lagrangian.\foot{We are skating over several technical issues at
this point. One should really introduce the entire panoply of
ghosts-for-ghosts. Moreover, there are Jacobian factors from the
change of variables. We expect that these all cancel, but have not
checked the details. }
 The result is a total derivative, leading
to a theory on the boundary of the form
\eqn\bdrythry{ \int_{\IR_t \times \Sigma_{D-2}} dt ({\p \over \p
t}
 \Phi_{p})\wedge d \Phi_{D-p-3} }
(There are several different versions of \bdrythry\
differing by various
integrations by parts.) We see that we should identify $\Phi_p$
and $d\Phi_{D-p-3}$ as conjugate variables in the sense of
Hamiltonian dynamics. The global nature of the gauge group is
therefore the global topology of the phase space, and will affect
the values that can be taken by coordinates and momenta.
Canonical quantization of the theory based on the action
\bdrythry\ leads to the Hilbert space of the theory of a $p$ form
gauge field (which is the same as the theory of a $D-p-3$ form
gauge field).

We can generalize the theory \bftheory\ by adding a surface term
to the action.  This term can depend on the metric on the
boundary. For example, we can add to the action
 \eqn\boundint{{i \over 4 g^2} \int_{\IR_t \times \Sigma_{D-2}}
 A_{D-p-2} \wedge {}^*A_{D-p-2}}
where ${}^*A_{D-p-2}$ is the dual in the boundary.  The
$i=\sqrt{-1}$ in front of \boundint\ shows that we view the time
direction $\IR_t$ as having Lorentzian signature. Now \bdrycond\
is replaced with
 \eqn\bdrycond{
\int_{\IR_t \times \Sigma_{D-2}} ({k \over 2\pi} A_{p+1} -{1 \over
2g^2} {}^*A_{D-p-2})\delta A_{D-p-2}=0. }
Free boundary conditions lead in this case to the condition
\eqn\sdcon{
{k
g^2 \over \pi}A_{p+1} ={}^*A_{D-p-2} ~.
}

We can integrate over $A_{p+1}$ in the bulk to find a delta
functional of $dA_{D-p-2}$ which means that $A_{D-p-2}$ is a flat
connection. Therefore, the bulk action \bftheory\ vanishes and
the boundary action \boundint\ can be written as
 \eqn\boundintp{{i\over 4g^2} \int_{\IR_t \times \Sigma_{D-2}}
 d\Phi_{D-p-3} \wedge {}^*d\Phi_{D-p-3}}
This is the standard theory of a $D-p-3$ form gauge field which
can be dualized to the theory of a massless $p$ form gauge field.

What is the difference between the theory based on \bdrythry\ and
the theory based on \boundintp?  The Hilbert space of the boundary
theory is that of a $p$ form gauge field in both cases, but the
Hamiltonian which acts on this Hilbert space is different in the
two theories.  The latter depends on the details of the boundary
interactions.

The case $D=4n+3$, $p=2n$ is particularly interesting.  Since
$p+1=D-p-2$ the two different gauge fields in \bftheory\ are of
the same degree and we can consider two linear combinations of
them $A_{2n+1}$ and $C_{2n+1}$ such that the action \bftheory\ can
be written as a difference of two terms $\int A_{2n+1}dA_{2n+1} -
C_{2n+1}dC_{2n+1}$. The spectrum of the theory on the boundary is
that of a $2n$ form gauge field. It splits into two contributions.
The selfdual part originates from $A$ and the  antiselfdual part
originates from $C$.  Therefore, we can consider the theory which
is based on $A$ only.  The spectrum on the boundary of this
theory is a chiral $2n$ form gauge field.

This discussion suggests a way to define an action for a chiral
gauge field on a $4n+2$ dimensional space $N_{4n+2}$.
\foot{The following discussion is closely
related to Witten's analysis of selfdual fields in
\refs{\imwitten,\duality}. Related ideas have also been
considered by S. Shatashvili.}
%%%%
%%%%
All we need
to do is consider the Chern-Simons action of a $2n+1$ dimensional
gauge fields in a manifold $M_{4n+3}$ such that $\partial
M_{4n+3} = N_{4n+2}$.  By adding various boundary interactions we
can study different Hamiltonians acting on the Hilbert space of
the chiral gauge field.

More explicitly, consider $N_{4n+2}=\IR_t\times \Sigma_{4n+1}$
with Lorentzian signature and study the action
 \eqn\chiact{\eqalign{
 S(A,B)=&-{ik\over 2\pi}\int_{M_{4n+3}}A_{2n+1}\wedge dA_{2n+1}\cr
 &+ {ik\over 4\pi}\int_{N_{4n+2}}(A_{2n+1}-B_{2n+1})\wedge{}^*
 (A_{2n+1}-B_{2n+1}) +2 A_{2n+1}\wedge B_{2n+1}\cr
 =&S(A,B=0)+ {ik\over 4\pi}\int_{N_{4n+2}}(A+{}^*A)\wedge
 (B-{}^*B) + B\wedge {}^*B \cr
 =&S(A,B=0)+ {ik\over 4\pi}\int_{N_{4n+2}}A^{(+)}\wedge
 B^{(-)} + B\wedge {}^*B  }}
where $B_{2n+1}$ is a nondynamical $2n+1$ form background field
(source) on $N_{4n+2}$.  Its anti-selfdual part $B^{(-)}$ couples
to the selfdual part of the dynamical field $A^{(+)}$.  The
boundary variation including the surface term from the bulk is
$[(A-B)-{}^*(A-B)]\delta A=0$ which for free boundary conditions
sets $A-B$ to a selfdual field.  Performing the functional
integral over $A$ the answer is $\exp[\Gamma(B)]$ where
$\Gamma(B)$ is a functional of $B$ . Since under the gauge
transformation which does not vanish on $N_{4n+2}$ and acts also
on the sources $\delta A=\delta B = d\lambda$ the action
transforms as $\delta S(A,B)= {ik\over 2\pi}\int_{N_{4n+2}}
d\lambda \wedge B $,
 \eqn\gammavar{\delta\Gamma(B)= {ik\over 2\pi}\int_{N_{4n+2}}
 d\lambda \wedge B }
Therefore, $\Gamma$ is a sum of a gauge invariant term and a
Chern-Simons term for $B$ on $M_{4n+3}$.

\subsec{Examples}

Let us flesh out this abstract discussion with some particular examples.

\item{1.} $D=2$ and $p=0$.
We now have action $S=k\int A d\wp $ where $A$ is a 1-form and
$\wp$ is a $0$-form. Let us suppose that the gauge group is such
that we  identify $A \sim A + \omega$ where $\omega$ is a closed
1-form with $2\pi Z$ periods. Let us work on the strip $[-L,L]\times R_t$. If we
integrate over $\wp(x,t)$ for $x\in (-L,L)$ then $A=dq$, and
substituting back into the action we get the basic action of
quantum mechanics
\eqn\qmact{
S = k \int dt q_1 \dot \wp_1 - k \int dt q_2 \dot \wp_2
}
where $q_1 = q(x=L,t), \wp_1 = \wp(x=L,t), \wp_2= \wp(x=-L,t),
q_2 = q(x=-L,t)$.
If we insert a ``Wilson line''
$\exp[ i e \int A_t dt ]$ then we can shift it away by
a ``singular gauge transformation''
\eqn\singgt{
\wp \to \wp - {e\over k} \theta(x)
}
Related to this, if we declare the gauge group of $\wp$ to
be $\wp \sim \wp+ Z$ then $\wp \to \wp+ N\theta(x)$ is a gauge
transformation which shifts the charge of
 a Wilson line by $q \to q + Nk $.
 Our choice of gauge group for $A$ means
$q\sim q+ 2\pi Z$. So, if we also identify $\wp\sim \wp + Z$
then we are quantizing the compact phase space $S^1 \times S^1$.
Then the momentum is indeed valued in $Z_k$.
On the other hand, we could choose not to identify $\wp$. Then
we are quantizing the phase space $S^1_q \times R_\wp$. Now
the momenta are indeed quantized, but can take any integral
value.

\item{2.}   $D=3$ and $p=0$. Since $D=4n+3$ and $p=2n$ here we
can have a single gauge field $A_1$.  This is the famous example
of 3d Abelian Chern-Simons theory, which is holographically dual
to the level $k$ $U(1)$ chiral algebra.  The Wilson loop operators
on the cylinder create holonomy $n/k$ for the gauge field. The
singular gauge transformations shift $n\to n+k$. These act
nontrivially on the Hilbert space associated with the boundary
theory, amounting to an insertion of the extending chiral field
$e^{i \sqrt{2k} \phi}$ of charge $k$ with respect to the current.
The different chiral sectors are labelled by $n \mod k$, and the
symmetry of the fusion rules is $Z_k$. Nevertheless, the
conformal field theory has a $U(1)$ symmetry.

\item{3.}  $D=5, p=1$. We have two 2-forms $B^{(1)}, B^{(2)}$. The
boundary theory has a $U(1)$ photon, $B^{(1)}=d A$, while
$B^{(2)}$ gives the dual photon.  The action \bdrythry\ for this
case was studied in \schse.

\item{4.} $D=7, p=2$. This example is similar to (2) and can have a
single gauge field $A_4$.

\appendix{B}{U(N) vs SU(N) in the AdS/CFT correspondence}

String theory on  $AdS_5 \times S^5$ is related to $U(N)$ Yang
Mills theory \review .
The $U(1)$ part of this $U(N)$ theory is free.
In \wittenhol\ and \aw , it was emphasized that bulk gravity in
AdS is dual to the $SU(N)$ part of the gauge theory.
It is clear from the arguments in \refs{\wittenhol,\aw} that the $U(1)$
degree of freedom cannot live in the bulk of AdS.

The theories in examples 2,3,4 in appendix A   appear in the low
energy description of string theory on AdS spaces. The
 singleton degrees of freedom play an important
role in determining the precise structure of the gauge
group in the holographically dual theory.
%In particular, the
%boundary conditions we impose at the boundary of $AdS$ will
%determine whether we include or we do not include the
%$U(1)$ degree of freedom.
As explained in appendix A, a Chern-Simons theory with
a boundary leads to a degree of freedom living at the boundary.
In the case of $AdS_5$ we have a Chern-Simons action in the bulk
for the  RR and NS  2-form potentials. These lead to a single
$U(1)$ gauge field on the boundary, which arises from the
``gauge freedom'' of the 2-form $B^{NS}$. (The gauge field
arising from $B^{RR}$ is the magnetic dual, see \sdcon. )

Let us consider the Wilson line in representation
$R$ along a closed curve $C$
in the boundary theory,  $W_{R,C}:=Tr_R Pexp \int_C {\bf A} $,
Here ${\bf A}$ is the $U(N)$ gauge field. It is an
$N\times N$ antihermitian matrix-valued 1-form.
Recall that an irreducible representation $R$ of $U(N)$ is
 labelled by a pair $(q,\lambda)$ where
$q$ is the integral charge specifying the representation of the
$U(1)$, and $\lambda$ is an irrep of $SU(N)$ such that
$q\mod N$ is the $N$-ality of $\lambda$.

In the AdS/CFT
correspondence the computation of the
expectation values of $W_{R,C}$ in the boundary theory
is replaced by the string theory correlators of Wilson surfaces
whose worldsheets $\Sigma$ end on $C$. Such worldsheets couple
to $B^{NS}$ as $\exp[i \int_{\Sigma} B^{NS}]$ and therefore
depend on the singleton gauge field $A$ in the same way as
a Wilson line for the $U(1)$ subgroup of charge $q=1$.

%%%%
%%%% minor changes here and below
%%%%
When $C$ is a contractible cycle in the boundary we are
computing the Wilson line associated to the creation/annihilation
of a quark/antiquark pair. This was computed via AdS/CFT in
\refs{\maldaloop,\reyloop}. On the other hand, if the boundary manifold
is of the form $S^1\times M_3$ then we can instead consider a Wilson
loop for $C= S^1 \times P$ where $P\in M_3$ is a point. Expectation
values of
\eqn\baryvtx{
\CW(P_1, \dots, P_N)  := {\rm Tr}_N\biggl( Pexp \int_{S^1\times P_1} {\bf A} \biggr)\cdots
{\rm Tr}_N\biggl( Pexp \int_{S^1\times P_N} {\bf A} \biggr)
}
where $P_i$,   $ i=1,..,N$,  are $N$ points on $M_3$,
measures the coupling to a ``baryon'' made out of the antisymmetric
combination of $N$ external quarks.
In \baryons\ and \aw\ it was argued that the existence,
in the gauge theory, of a
``baryon vertex,'' namely  a gauge invariant coupling of $N$ external
quarks  with nonzero expectation value,
 implies that the bulk physics is described by the $SU(N)$
part of the gauge theory. We now  can see that if we include the
singleton mode that lives at the boundary we can also account for the
$U(1)$ degree of freedom.

At first sight the baryon vertex seems to preclude the existence
 of a $U(1)$ degree of freedom
living at the boundary. One  argument
for this is that in the path  integral evaluation of
\baryvtx\ the integral over the zeromode of
$\int_{S^1\times P}Tr{\bf A}$ would set expectation values
of \baryvtx\ and its products to zero.
\foot{This issue did not arise in the computation of
\refs{\maldaloop,\reyloop}, because, for a contractible $C$,  there is
no zeromode to integrate over.}
 The apparent contradiction is resolved as follows.
\foot{A second, essentially equivalent, resolution proceeds by
studying the $U(1)$ gauge invariance of an operator
related to \baryvtx\ but involving open Wilson lines.}
For definiteness, take $M_3= S^3$ and fill it in with
the disk $D^4$.
The gravity dual for expectation values of \baryvtx\ involves
an insertion in the bulk of a D5-brane wrapped on
$Q \times S^1 \times S^5$, where $Q\in D^4$ is a point \baryons.
As explained in \baryons\ charge conservation forces us
to attach $N$ Wilson surfaces coupling to $B^{NS}$.
However, we must also include a ``Dirac string'' singularity
of the field $B^{RR}$. This ``Dirac string'' is
actually  a two dimensional surface
 in the five-dimensional manifold $D^4 \times S^1$, and a
 7-dimensional manifold in the full spacetime, but we will
refer to it as a ``Dirac string.''
While the Dirac string for $B^{RR}$ has no
physical effect in the bulk, it {\it does}
have a physical effect when it intersects the boundary
on $S^1 \times P_0$, where $P_0$ is a point on $S^3$. Indeed, it
acts as a source of charge $-N$ for the singleton mode
holographically dual to $B^{NS}$.
The reason for this is the following. The RR field around this
Dirac surface is such that $\int_{S^2} B^{RR} = 2 \pi $, where
$S^2$ is the sphere linking the Dirac surface in five dimensions.
As we explained above $B^{NS} $ on the boundary is the field strength
of the singleton gauge field, while $B^{RR}$ is the field
strength of the electric-magnetic dual singleton field.
%%%
Then, one can either use singular gauge transformations in  the
BF theory $S= N \int d B^{RR} \wedge B^{NS}$, or, alternatively,
invoke boundary conditions
\sdcon\ to  conclude that there is an electric field for the
singleton field of strength $-N$.

The conclusion from the above reasoning is that
{\it the holographic
dual of the insertion of a wrapped $D5$ brane is the
expectation value of }
\eqn\modifyw{
\exp\bigl[ -\int_{C\times P_0} {\Tr} {\bf A}\bigr] \CW(P_1, \dots, P_N).
}
 That is, in terms of the gauge theory,
we have $N$ U(N) quarks inserted where the fundamental
strings intersect the
boundary, and we have a Wilson line  which couples only to the $U(1)$
with charge $-N$ inserted at  $P_0\in S^3$ (and
winding along the $S^1$). Similarly, in cases where we have  flux of
$H_{RR}$ on $S^3$ we will have Dirac surfaces intersecting
$S^3$ at points. Again, each Dirac surface intersecting the boundary
leads to the insertion of a Wilson loop for the $U(1)$ with charge $N$,
in the appropriate duality frame.

There is no obvious reason for \modifyw\  to vanish
from integration over a zeromode. Moreover, the result
of the computation will depend on the point $P_0$ where the Dirac string
intersects the boundary. This dependence   depends on the
Hamiltonian for the singleton, which in turn we can view as arising
from a choice in the boundary conditions at infinity.

\ifig\baryonads{We consider a fixed time slice of the $AdS$ geometry.
Here we see $N$ strings  ending on a baryon vertex. A Dirac
string, indicated by the doted line,
 emanates from the wrapped $D5$ brane. At the point where this string
crosses the boundary of AdS there is an operator in the $U(1)$ theory
with charge $-N$ inserted.
}
{\epsfxsize1.5in\epsfbox{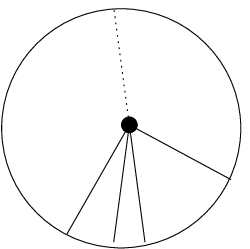}}

Even though the $U(1)$ singleton degree of freedom has no
consequence for bulk physics in $AdS$ it is important to
understand its origin in order to compare field theory answers
with gravity answers. In particular, if we calculate Wilson loop
expectation values in the gravity picture we need to understand,
and state precisely, the boundary conditions for the $B$-fields
on the boundary of $AdS$. In particular, if we do the computation
by including the asymptotically flat region, then we are
automatically including the $U(1)$ degree of freedom, see for
example \refs{\BilalTY,\mv} for computations where it is
automatically included.  So when we take the decoupling limit, the
finite answer that we will get will include the $U(1)$ degrees of
freedom. Similar care should be exercised when we do computations
of anomalies, etc, in the bulk theory.

In the discussion of \aw\ a prominent role was
played by
\eqn\defalpha{
\alpha = \int_{D_2} B_{NS}
}
in cases where the five dimensional geometry was $D_2 \times S^3$ or
$D_2 \times R^3$. It is important to understand the field
theory interpretation of this variable.
Consider a Wilson line  $ W = Tr( P e^{\int_{S^1} {\bf A}}) $, we
define $\alpha$ through $W = |W|e^{i \alpha}$.
When
we introduce this Wilson loop operator in the path integral, via
the AdS prescription, we indeed get a phase equal to \defalpha .
The phase
$e^{i\alpha}$ depends   on both the $SU(N) $ degrees of freedom and on
the $U(1)$ degrees of freedom. If we are considering an
infinite volume system, the field theory on
 $S^1 \times R^3$ for example, then $\alpha$ is fixed
as a vev.
 In \aw\ the behaviour of various
quantities under the $SU(N)$ dependence of $\alpha$ was studied.
In particular, it was argued that $\alpha \to \alpha + 2\pi/N$
should be a symmetry, but $\alpha \to \alpha + \beta$ should not
be a symmetry for arbitrary $\beta$.
If we perform a gauge transformation on the
$B$ field that does not vanish at infinity we see that
we change $\alpha \to \alpha + \beta$, and this  is equivalent to
introducing a Wilson line for the singleton.
In \aw\ these  gauge transformations were not allowed
so that the singleton mode was effectively
frozen to zero,  so that
 only the dependence of the physical answers
on the SU(N) dependence of $\alpha$ were studied.

It is amusing to note that in the Klebanov-Witten
\klebanovwitten\ solution there is only one obvious
 set of $B$ fields
with Chern Simons terms, the same we had above, and therefore only one
$U(1)$. From the field theory point of view we could imagine
starting with a $U(N)\times U(N)$ theory so that we would naively
expect two $U(1)$'s. The relative $U(1)$ is truly decoupled in the
IR by the renormalization group flow
while the overall $U(1)$ is the one we still see living at the
boundary and coming from the $B-$fields in $AdS$. Baryons
 that come from D3 branes in the bulk \gk\ are not charged
under this overall $U(1)$. The question of whether the relative
$U(1)$ appears or not seems hard to answer.

What we have said above regarding the singleton mode in the case
of $AdS_5\times S^5$, can also be generalized to other dimensions.
In the case  of $AdS_7\times S^4$
 there is a three form potential
$C_3$  in seven dimensions with a
coupling $ { N \over 4 \pi }  \int_{AdS_7} C_3 \times dC_3$. This
$C_3$ leads to a self dual two form on the boundary.
In the $AdS_3$ case, gauge fields in $AdS_3$, with Chern Simons
couplings lead to chiral scalar fields on the boundary.
In the $AdS_4\times S^7$
case all modes living on the boundary are expected
to be scalar fields and they should
be modes with $SO(8)$ quantum numbers.
It would be interesting to see them.

\listrefs

\bye